\documentclass[aps,pre]{revtex4-1}
\usepackage{color,bm,epsfig,graphics,amssymb,amsmath,subeqnarray,setspace,graphicx,amsthm,epstopdf,subfigure,color}

\def\b{\mathbf}\def\e{\epsilon}

\begin{document}
\title{Jet propulsion without inertia}
\author{Saverio E. Spagnolie}
\email{sespagnolie@ucsd.edu}
\author{Eric Lauga}
\email{elauga@ucsd.edu}
\affiliation{Department of Mechanical and Aerospace Engineering, University of California San Diego, 9500 Gilman Drive, La Jolla CA 92093-0411.}
\date{\today}

\begin{abstract}
A body immersed in a highly viscous fluid can locomote by drawing in and expelling fluid through pores at its surface. We consider  this  mechanism of jet propulsion without inertia in the case of spheroidal bodies, and derive both the swimming velocity and the hydrodynamic efficiency. Elementary examples are presented, and exact axisymmetric solutions for spherical, prolate spheroidal, and oblate spheroidal body shapes are provided. In each case, entirely and partially porous (i.e. jetting) surfaces are considered, and the optimal jetting flow profiles at the surface for maximizing the hydrodynamic efficiency are determined computationally. The maximal efficiency which may be achieved by a sphere using such jet propulsion is $12.5\%$, a significant improvement upon traditional flagella-based means of locomotion at zero Reynolds number. Unlike other swimming mechanisms which rely on the presentation of a small cross section in the direction of motion,  the efficiency of a jetting body at low Reynolds number increases as the body becomes more oblate, and limits to approximately $162\%$ in the case of a flat plate swimming along its axis of symmetry. Our results are discussed in the light of slime extrusion mechanisms occurring in many cyanobacteria.
\end{abstract}

\maketitle

\section{Introduction}

Locomotion at the micron scale, or in highly viscous fluids, is constrained by the dominance of viscous dissipation over inertial effects. Hence, the strategies utilized by microorganisms and engineered swimming devices must differ from more familiar locomotive mechanisms such as the flapping of wings \cite{wbb75,Purcell77,Childress81,Vogel94}. Instead of imparting momentum into a fluid wake, which is not possible at zero Reynolds number, bodies generally exploit drag anisotropy in order to propel themselves through the fluid. In particular, the undulation of flagella and cilia are the most well-studied means of swimming at low Reynolds numbers in nature. As shown in classical explorations of flagellar locomotion \cite{Lighthill75,bw77}, the maximum theoretical hydrodynamic efficiency which may be achieved by flagellar propulsion is approximately $8\,\%$, and is generally closer to $2\,\%$ in real cells \cite{Lighthill75,lp09,sl10}. For locomotion using cyclic tangential surface distortions, the efficiency has a theoretical bound of $75\%$ in the case of a spherical body, a dramatic improvement upon that granted by the use of flagella
 \cite{ss96}.

Flagellar locomotion is not the only means of locomotion at low Reynolds numbers and many authors have considered alternatives, either to help explain biological phenomena, or to suggest designs for synthetic locomotor systems on small scales. Examples of swimming bodies which deform in a manner which breaks the important time-reversal symmetry (a constraint known as the Scallop theorem and without which no locomotion is possible in the absence of inertia) were presented by Purcell \cite{Purcell77}. These systems include the motion of a three-link swimmer \cite{Purcell77,bks03,th07,ar08} and a treadmilling torus \cite{Purcell77,Taylor52,kts05,tsk07,lk08}. Recently, Leshansky et al. \cite{lk07} have considered the remarkably efficient locomotion of an elongated treadmilling body, which can propel itself at nearly the same velocity of the surface motion. Other simple means of propulsion proposed for locomotion at zero Reynolds number include a deformable two-dimensional loop \cite{ag04}, systems of two \cite{ako05,opw08}, three \cite{najafi05,znm09} or $N$ spheres \cite{Felderhof06}, flapping near deformable interfaces \cite{ty08}, a rehinging swimmer \cite{Spagnolie09}, and a jellyfish-inspired bilayer vesicle \cite{ESL10}. For a more complete list of references we refer the reader to Ref.~\cite{lp09}.

\begin{figure}[htbp]
\begin{center}
\includegraphics[width=4in]{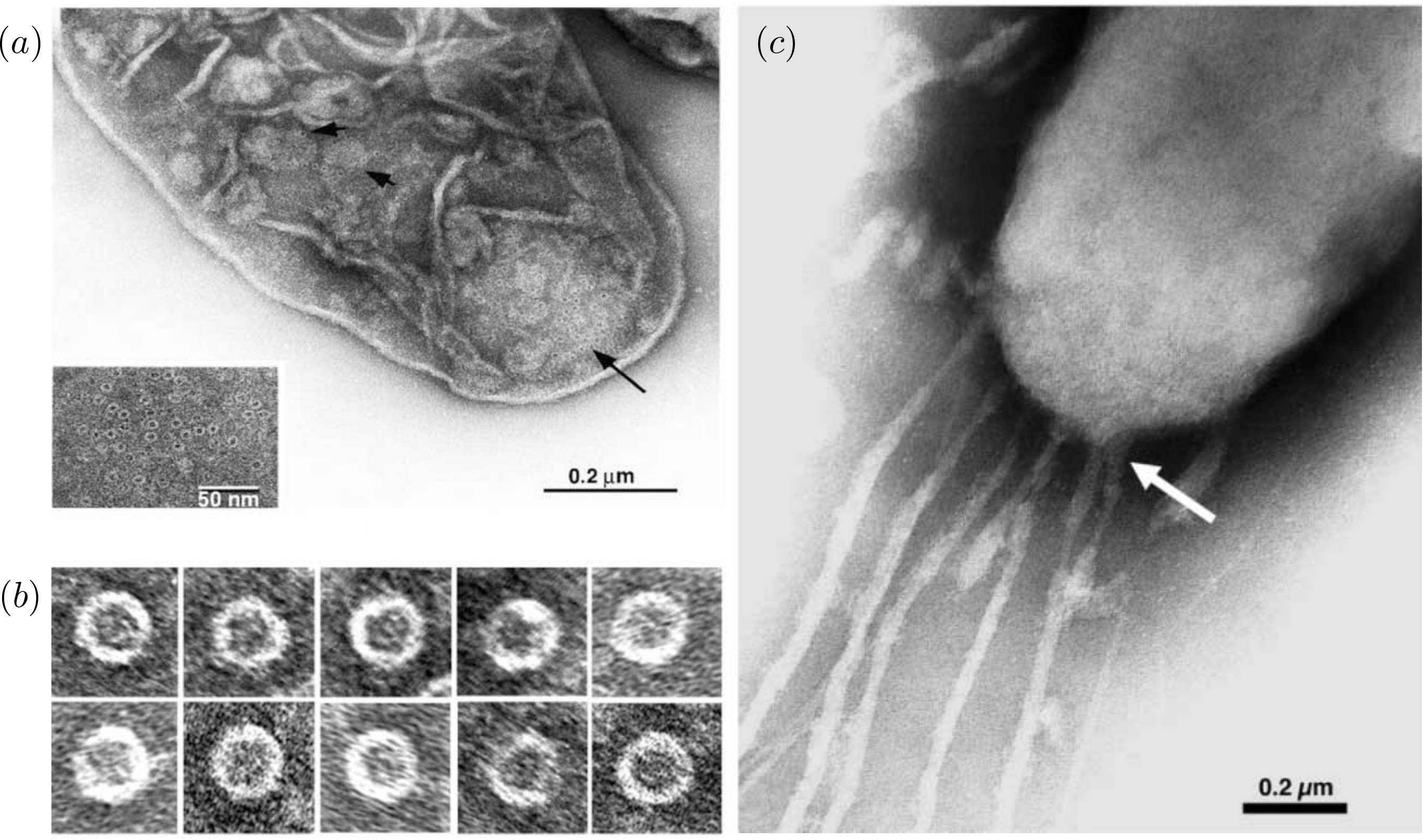}
\caption{Reprinted from Wolgemuth et al. \cite{whko02} with permission from Elsevier. (a) Negatively stained electron micrograph of an isolated {\em M. xanthus} cell envelope showing multiple ring-like structures located predominantly at the poles of the cell. The inset shows a higher magnification of the nozzle array in the region indicated by the long arrow. (b) A gallery of electron micrographs of negatively stained isolated nozzles. Each cylindrically symmetric nozzle has an outer diameter of 14 nm. (c) Electron micrograph of a gliding cell. At higher magnification, it can be seen that the slime trails are composed of several slime bands, which are secreted from the sites at the cell pole, where the nozzles are located (large arrow).}
\label{Figure1}
\end{center}
\end{figure}

Recently, it has been observed that certain types of bacteria, such as myxobacteria and cyanobacteria (blue-green algae) secrete mucilage through nozzle-like organelles while gliding along a substrate (see Fig.~\ref{Figure1}) \cite{Walsby68,gc82,hb98,Hoiczyk00,whko02,wo04}. This so-called ``slime extrusion'' has been theorized as a primary propulsion mechanism for adventurous motility in such organisms as {\em M. xanthus}. The slime is a polyelectrolyte gel, and has been modeled after snail slime. Wolgemuth et al. \cite{whko02} have shown that the osmotic expansion of the slime from the nozzle generates a sufficient force to propel the organism. As the slime exits the nozzle, it adheres to the substratum, and further slime extrusion produces a thrust. Though the primary activation is located near the poles of the elongated body, the entire surface is covered with nozzles. Meanwhile, the motility mechanism employed by the motile marine cyanobacteria {\em Synechococcus} is still an open problem in biophysics \cite{wwfvw85,esbm96,ss96,Brahamsha99}. {\em Synechococcus} is known to swim absent the presence of a substrate, without changing shape, and without any observable external organelles, a puzzle to which we shall return.

Inspired by the gliding locomotion of the cyanobacteria, we consider in this paper the following questions: Can the extrusion of a Newtonian fluid be used as a propulsive mechanism absent a solid surface? And if so, how efficient would such a swimmer be? Herein we study the swimming motion which may be achieved by the placement of surface nozzles upon a body surface which act to both draw in and expel fluid, as illustrated in Fig.~\ref{Figure2}. We refer to this locomotion as {\em extrusion swimming}, or {\em non-inertial jet propulsion}. The swimming mechanism is not unlike the more familiar jet propulsion at higher Reynolds numbers, but with distinctly different flow structure and resultant (non-local) fluid-body interactions. In classical high Reynolds number jet propulsion, a body propels itself by imparting momentum onto the fluid opposite the direction of motion (as in the swimming of jellyfish) \cite{Weihs77,Daniel84,Vogel94,dg88,lt04,dg05,Mohseni06,Dabiri08}. The low Reynolds number analogue studied here propels itself instead by taking advantage of the viscous stresses induced by the jet motion on its surface.

The fluid being expelled from the body is assumed to be Newtonian and identical to the surrounding fluid, and the body is assumed to be well-separated from any surfaces. The swimming velocity and hydrodynamic efficiency of such a body are first derived formally using the Lorentz reciprocal identity. We find that the maximal hydrodynamic efficiency which may be achieved by a sphere using such jets is $12.5\%$, a significant improvement upon most other means of swimming at zero Reynolds number. Moreover, unlike many other swimming mechanisms which rely on the presentation of a small surface area in the direction of motion, we show that the  efficiency increases as the body becomes more oblate, and limits to approximately $162\%$ in the case of a flat plate moving along its axis of symmetry. 

The manuscript is organized as follows. In \S II, the swimming velocity of an arbitrary body is shown to be dependent upon the fluid stress in a dual (but simpler) resistance problem, and we present elementary examples for certain swimming spheroids and a rotating ``viscous pinwheel.'' The hydrodynamic efficiency of locomotion is also defined. In \S III, we consider the behavior and efficiency of a spherical body, for the case of an entirely porous (i.e. jetting) surface, and then for a partially porous surface. We also determine the optimal jetting flow profile at the surface for maximizing the swimming efficiency. In \S IV and \S V we consider prolate and oblate spheroidal bodies, respectively. Entirely and partially porous surfaces are considered, and jetting profile optimization is performed. We conclude by applying the theoretical results to the organism {\em Synechococcus}, which swims by a yet-unknown mechanism, and show that while slime extrusion is important for gliding motility in related cyanobacteria, extrusion of a Newtonian fluid cannot not properly account for its observed swimming speed.

\section{General expressions for the swimming velocity and efficiency}

\subsection{Description of fluid-jetting bodies and fluid-body interactions}
\begin{figure}[t]
\begin{center}
\includegraphics[width=4.5in]{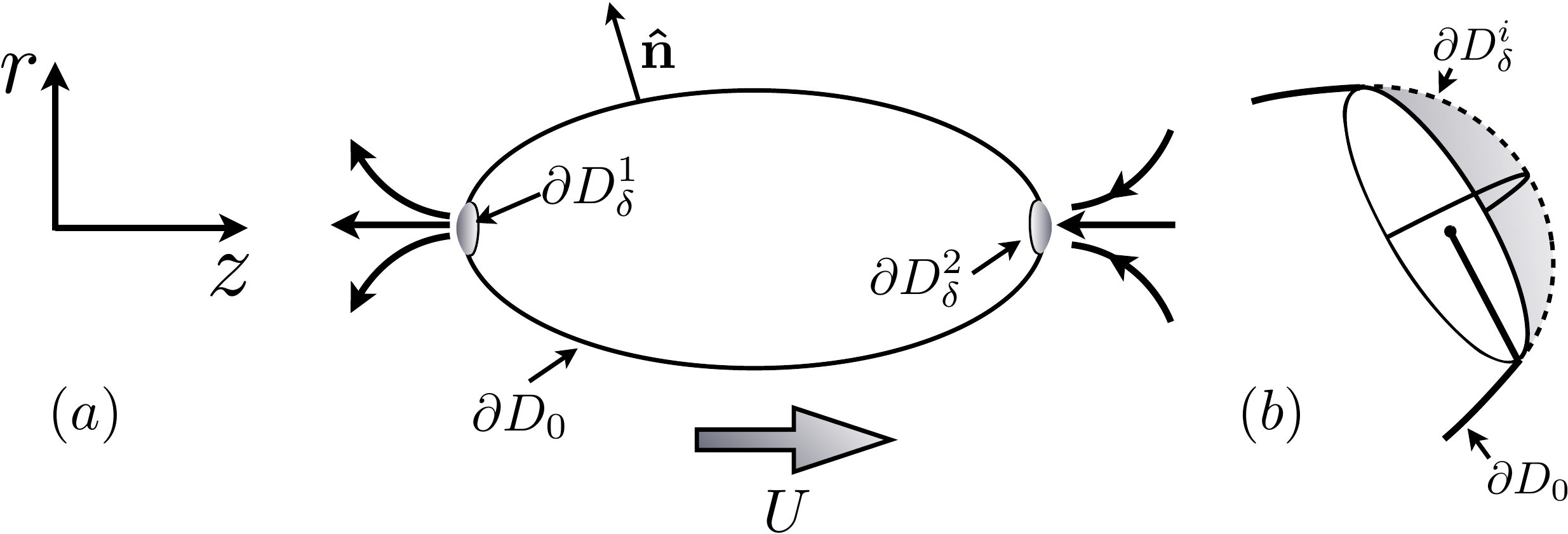}
\caption{(a) A fluid-jetting body. The surface is composed of a solid part, $\partial D_0$, and the $m$ porous surfaces which allow for inflow and outflow, $\partial D_\delta ^i$ for $i=1,2,...,m$. The entire surface is denoted by $\partial D=\partial D_0 \displaystyle\cup_{i=1}^m\partial D_\delta^i$. Here the body draws in and expels fluid to the left, and swims to the right with velocity $U$. We denote by $\b{\hat{n}}$ the unit normal vector pointing into the fluid. (b) The surface $\partial D_\delta^i$ corresponds to the $i^{th}$ fluid jet, or porous surface.}
\label{Figure2}
\end{center}
\end{figure}

We begin our consideration by calculating the swimming velocity of a jetting body of arbitrary shape. A fluid-jetting body is illustrated in Fig.~\ref{Figure2}. The surface is composed of a solid part, $\partial D_0$, and the $m$ porous surfaces which allow for inflow and outflow, $\partial D_\delta ^i$ for $i=1,2,...,m$. A no-slip condition is assumed to hold on the surface $\partial D_0$, and the porous flow through the surfaces $\partial D_\delta^i$ is assumed to be driven by internal mechanisms in the direction normal to the surface at each point, either into or out of the body. The entire surface is denoted by $\partial D=\partial D_0 \displaystyle\cup_{i=1}^m\partial D_\delta^i$. 

The equations for incompressible fluid motion at zero Reynolds number are the Stokes equations,
\begin{gather}
\nabla \cdot \bm{\sigma}=-\nabla p + \mu\,\Delta \b{u}=\b{0},\label{E:Stokes}\\ 
\nabla\cdot \b{u}=0,\label{E:Incompressibility}
\end{gather}
where $\bm{\sigma}=-p\b{I}+2\mu \b{E}$ is the Newtonian fluid stress, $p$ is the dynamic pressure, $\b{u}$ is the fluid velocity, $\b{E}=(\nabla \b{u}+\nabla \b{u}^T)/2$ is the symmetric rate-of-strain tensor, and $\mu$ is the shear viscosity. The boundary conditions are decomposed into rigid body motion and a fluid extrusion (or jetting flow) component,
\begin{align}
\b{u(x)}&=\b{U}+\b{\Omega}\times \b{x}+\phi(\b{x})\b{\hat{n}}(\b{x})\,\,\,\,\,\,\,\,(\b{x}\in \partial D),\label{E:poreBC}\\
\b{u(x)}&\rightarrow \b{0},\,\,\,\,\,p\rightarrow 0\,\,\,\,\,\,\,(\b{x}\rightarrow \infty).\label{E:BCs}
\end{align}
The jetting-flow profile $\phi(\b{x})$ has support only at the porous surfaces, $\b{x}\in \cup_{i=1}^m \partial D_\delta^i$, or $\phi(\b{x})=0$ for $\b{x}\in \partial D_0$. We denote by $\b{\hat{n}}(\b{x})$ the unit normal vector at the point $\b{x}$ pointing into the fluid, and $\b{U}$ and $\b{\Omega}$ are the instantaneous rigid body velocity and rotation rate resulting from the fluid extrusion. We define the fluid-jetting fluxes (i.e. flow rates) $q_i$ via
\begin{gather}
\int_{\partial D_\delta^i}\phi(\b{x})\,dS=q_i,
\end{gather}
with $dS$ the surface area element. The internal volume is conserved if the inward and outward fluxes balance, 
\begin{gather}
\displaystyle \int_{\partial D}\b{\hat{n}}\cdot \left(\phi(\b{x})\,\b{\hat{n}}\right)\,dS=\sum_{i=1}^m\,q_i=0.
\end{gather}
Finally, to close this ``swimming problem'' we require that the body be force and torque free, i.e.
\begin{gather}
\int_{\partial D} \bm{\sigma}\cdot \b{\hat{n}}\,dS=\bm{0},\,\,\,\,\,\,\int_{\partial D} \bm{x}\times \left(\bm{\sigma}\cdot \b{\hat{n}}\right)\,dS=\bm{0}.
\end{gather}

\subsection{General expression for the swimming velocity}

An identity attributed to Lorentz may be used to deduce the rigid body motion resulting from the porous jetting flow on the body surface. This approach was used by Brenner to compute the drag on a body in an arbitrary flow field \cite{Brenner64}, and by Stone and Samuel to study the swimming of finite size bodies \cite{ss96}. The Lorenz reciprocal theorem is stated as
\begin{gather}
\int_{\partial D} \b{\tilde{u}}\cdot \left(\bm{\sigma} \cdot \b{\hat{n}}\right)\,dS=\int_{\partial D} \b{u}\cdot \left(\bm{\tilde{\sigma}} \cdot \b{\hat{n}}\right)\,dS,\label{E:Lorentz}
\end{gather}
where $(\b{u},\bm{\sigma})$ are the velocity and stress fields corresponding to the swimming problem described above, and $(\b{\tilde{u}}, \bm{\tilde{\sigma}})$ are the velocity and stress fields corresponding to a different system, but one which shares the same  instantaneous immersed boundary (see Refs.~\cite{hb65,kk91}). Let $(\b{\tilde{u}}, \bm{\tilde{\sigma}})$ be the solution to the Stokes equations for rigid body motion of the immersed body, so that the surface fluid velocity is $\b{\tilde{u}}=\b{\tilde{U}+\tilde{\Omega}\times x}$ and the corresponding net forces and torques are $\b{\tilde{F}}$ and $\b{\tilde{L}}$ respectively (the ``resistance problem''). Since the body is force and torque free in the swimming problem, the integral on the left hand side of Eq.~\eqref{E:Lorentz} vanishes. The remaining terms in the reciprocal identity are
\begin{gather}
\bm{0}=\int_{\partial D} \left(\b{U+\Omega \times x}+\phi(\b{x})\,\b{\hat{n}}\right)\cdot \left(\bm{\tilde{\sigma}} \cdot \b{\hat{n}}\right)\,dS.
\end{gather}
Since $\phi(\b{x})=0$ for $\b{x}\in \partial D_0$, we obtain
\begin{gather}
\b{U}\cdot\b{\tilde{F}}+\b{\Omega\cdot \tilde{L}}=-\int_{\partial D} \phi(\b{x})\left(\b{\hat{n}}\cdot \bm{\tilde{\sigma}} \cdot \b{\hat{n}}\right)\,dS=-\displaystyle \sum_{i=1}^m\int_{\partial D_\delta^i} \phi(\b{x})\left(\b{\hat{n}}\cdot \bm{\tilde{\sigma}} \cdot \b{\hat{n}}\right)\,dS.\label{E:MainSwimmingVelocity}
\end{gather}
In the event that a body is to translate or rotate with the greatest velocity for a given set of jet fluxes $q_i$, Eq.~\eqref{E:MainSwimmingVelocity} indicates that the jets should be placed precisely where the body experiences its largest normal component of traction in the corresponding resistance problem. Also of note, the consequences on the flow and swimming velocities from inward flowing jets are precisely opposite those of outward flowing jets. Unlike at higher Reynolds numbers where there is a distinct asymmetry in the flow fields set up by an outward flux and inward flux through a small opening (candles are blown out instead of sucked out!), at zero Reynolds number sources and sinks near a wall produce identical streamlines \cite{hb65}. We proceed to consider a number of simple examples which utilize Eq.~\eqref{E:MainSwimmingVelocity}.

\subsection{Example 1: A translating sphere}
As a first example, consider the simple case of a translating sphere of radius $a$. The corresponding resistance problem is well known: $\b{\tilde{F}}=6\pi\mu a \b{\tilde{U}}$, $\b{\tilde{L}}=\b{0}$, and $(\bm{\tilde{\sigma}} \cdot \b{\hat{n}})=3\mu/(2 a)\b{\tilde{U}}$ on the body surface \cite{hb65}. Generally, then, the swimming velocity of a jetting sphere may be written as
\begin{gather}
\b{U}=-\frac{1}{4\pi a^2}\int_{\partial D}\phi(\b{x})\b{\hat{n}}\,dS.\label{E:SphereU}
\end{gather}

The largest swimming velocity for a balanced inward/outward fluid flux is therefore achieved by placing two jets of equal and opposite strengths at the spherical poles along the axis of locomotion ($\b{x}_1$ and $\b{x}_2$),
\begin{gather}
\phi(\b{x})=\displaystyle q\left(\delta(\b{x},\b{x}_1)-\delta(\b{x},\b{x}_2)\right),\label{E:ExamplePhi}
\end{gather}
where $\delta(\b{x},\b{y})$ is the Dirac delta function with support at $\b{x}=\b{y}$, and $q=q_1=-q_2$. The corresponding swimming velocity is
\begin{gather}
\b{U}=-\frac{q}{2\pi a^2}\cdot
\end{gather}
While the flow profile $\phi(\b{x})$ stated above yields the largest swimming velocity, it does not present the most efficient means of moving the sphere through the fluid, as we shall explore in \S III. 

More generally, if the pores are small, $S_\delta^i \ll 1$ (with $S_\delta^i$ the surface area of the $i^{th}$ porous surface $\partial D_\delta^i$), then it is useful to expand the stress from the resistance problem linearly about the jet locations. For instance, near the $i^{th}$ porous jet we have
\begin{gather}
\b{\hat{n}}(\b{x})=\b{\hat{n}}(\b{x}_i)+O(S_\delta^i),\\
(\bm{\tilde{\sigma}} \cdot \b{\hat{n}})(\b{x})=(\bm{\tilde{\sigma} }\cdot \b{\hat{n}})(\b{x}_i)+O(S_\delta^i),
\end{gather}
and Eq.~\eqref{E:MainSwimmingVelocity} may be written as
\begin{gather}
\b{U}\cdot\b{\tilde{F}}+\b{\Omega\cdot \tilde{L}}=-\displaystyle \sum_{i=1}^m q_i \,\left(\b{\hat{n}}\cdot\bm{\tilde{\sigma}}\cdot \b{\hat{n}}\right)(\b{x}_i)+O(\displaystyle \max_i S_\delta^i).\label{E:mjets}
\end{gather}
The swimming velocity of a sphere with $m$ small fluid jets acting upon its surface is, from Eq.~\eqref{E:mjets},
\begin{gather}
\b{U}=-\frac{1}{4\pi a^2}\displaystyle \sum_{i=1}^m q_i \,\b{\hat{n}}(\b{x}_i)+O(\displaystyle \max_i S_\delta^i).
\end{gather}
A jetting sphere may therefore move in any direction spanned by the normal vectors at the jet locations by tuning the fluxes $q_i$, and the resulting motion will be a simple translation in that direction. Regardless of their distribution and strengths, jets acting normal to the body surface cannot, however, be used to generate rotations of a sphere. For the case of a rotating sphere of radius $a$, the resistance problem has $\b{\tilde{F}}=\bm{0}$,  $\bm{\tilde{L}}=8\pi\mu a^3\b{\tilde{\Omega}}$, but also $\left(\b{\hat{n}}\cdot \bm{\tilde{\sigma}} \cdot \b{\hat{n}}\right)=\bm{0}$, so that Eq.~\eqref{E:MainSwimmingVelocity} gives $\b{\Omega}=\b{0}$ regardless of the jetting flow profile $\phi(\b{x})$. In order to rotate using porous extrusion, the body must not be axisymmetric about the axis of rotation. 

\subsection{Example 2: A spheroid translating along its axis of symmetry}

The swimming velocity of a jetting spheroid translating along its axis of symmetry can also be determined with ease. Once again the corresponding resistance problem has a long history \cite{hb65}. Consider a prolate spheroid with major and minor axis lengths $2a$ and $2b$, with its major axis aligned with the $z$ axis. Setting $\b{\tilde{U}}=\tilde{U}\b{\hat{z}}$, we have $\b{\tilde{F}}=6\pi\mu R \b{\tilde{U}}$, $\b{\tilde{L}}=\b{0}$, and 
\begin{gather}
\bm{\tilde{\sigma}}\cdot \b{\hat{n}}=-\left(\frac{2\mu \,\tilde{U} \zeta}{c(\tau_0^2-\zeta^2)[(\tau_0^2+1)\coth^{-1}(\tau_0)-\tau_0]}\right) \b{\hat{n}}
\end{gather}
on the body surface, where $\zeta=\cos(\theta)$ ($\theta$ is the polar angle), $c=\sqrt{a^2-b^2}$, $\tau_0=a/c$, and
\begin{gather}
R=\frac{8c/3}{ \left(\tau_0^2+1\right) \log\left[\displaystyle\frac{\tau_0+1}{\tau_0-1}\right]-2\, \tau_0}\label{E:Rdef}
\end{gather}
(see \cite{hb65}). Equation~\eqref{E:MainSwimmingVelocity} then gives the swimming velocity; after some algebra we find that
\begin{gather}
U=-\frac{1}{2}\int_{-1}^1 G(\zeta)\phi(\b{\zeta})\,d\zeta,\quad 
G(\zeta)=\zeta\sqrt{\frac{\tau_0^2-\zeta^2}{\tau_0^2-1}}\cdot
\end{gather}
The function $G(\zeta)$ is monotonically increasing in $\zeta$ for all values of $\tau_0$ with $G(0)=0$ and $G(1)=1$. Confirming intuition, jets placed nearer to the poles contribute more significantly to the swimming velocity. Placing jets only at the poles, which expel and draw in fluid with fluxes $\pm q$, the swimming velocity (with $\b{U}=U\b{\hat{z}}$) is
\begin{gather}
U=-\frac{q}{2\pi b^2 }\cdot
\end{gather}
For $b=a$ we recover the spherical swimming velocity, and as the body becomes more slender the swimming velocity increases without bound (for fixed $q$). This is the largest velocity which may be achieved by a jetting prolate ellipsoid in the direction of its major axis for a given inward/outward flux, and as we will show in \S IV is again not the most efficient.

The swimming velocity of an oblate ellipsoid translating along the symmetric axis can be obtained by applying to the above the transformation $(c,\tau_0)\rightarrow (i\,c,-i\,\lambda_0)$, and reversing the definitions of $a$ and $b$ so that $\lambda_0=b/\sqrt{a^2-b^2}>0$ (see Ref.~\cite{hb65}). In this case the body swims with velocity
\begin{gather}
U=-\frac{1}{2}\int_{-1}^1 \zeta\sqrt{\frac{\lambda_0^2+\zeta^2}{\lambda_0^2+1}}\phi(\b{\zeta})\,d\zeta,
\end{gather}
which is again maximized by placing jets at the poles passing fluid with fluxes $\pm q$, giving $U=-q/(2\pi a^2)$. For fixed $q$, the swimming velocity decreases without bound as the presented surface area and hence fluid drag (due to the large no-slip surface area) increases.

\subsection{Example 3: A prolate spheroid translating along its minor axis}

As a third example, we determine the motion of a prolate spheroid translating along its minor axis. We consider the same prolate spheroid as in the previous example, and compute the swimming velocity $\b{U}=U\b{\hat{x}}$ when two jets (with fluxes $\pm q$) are placed at $\b{x}=\pm b\, \b{\hat{x}}$. The resistance problem has a tractable solution and representation using the singularity methods described by Chwang and Wu \cite{cw75}. In that work it was shown that a prolate spheroid translating along its minor axis with velocity $\b{\tilde{U}}=\tilde{U}\b{\hat{x}}$ generates flow and pressure fields given by
\begin{eqnarray}
\b{u}&=&\tilde{U}\b{\hat{x}}-\alpha B_{1}\b{\hat{x}}-\alpha x\left(\frac{1}{R_2}-\frac{1}{R_1}\right)\b{\hat{z}}-\alpha x r B_{3}\b{\hat{r}} +\nabla \left\{\beta x\left[\frac{z-c}{r^2}R_1-\frac{z+c}{r^2}R_2 +B_{1}\right]\right\},\\
p&=&2\mu \alpha \frac{x}{r^2}\left(\frac{z-c}{R_2}-\frac{z+c}{R_1}\right),
\end{eqnarray}
where $c=\sqrt{a^2-b^2}$, $r=\sqrt{x^2+y^2}$, $\b{\hat{r}}=(x \b{\hat{x}}+y \b{\hat{y}})/r$, $R_1=\sqrt{(z+c)^2+r^2}$, $R_2=\sqrt{(z-c)^2+r^2}$,
\begin{gather}
B_1=\log\left(\frac{R_2-(z-c)}{R_1-(z+c)}\right),\quad B_3=\frac{1}{r^2}\left(\frac{z+c}{R_1}-\frac{z-c}{R_2}\right),
\end{gather}
and finally,
\begin{gather}
\alpha=\frac{2\beta e^2}{1-e^2}=\tilde{U}e^2\left[e+(3e^2-1)\tanh^{-1}(e)\right]^{-1},
\end{gather}
with $e=c/a$ the eccentricity. From the flow and pressure fields the stress tensor $\bm{\tilde{\sigma}}$ may be computed without much difficulty. Integrating the stress over the body surface results in the expression
\begin{gather}
\b{\tilde{F}}=6\pi \mu C \tilde{U}\b{\hat{x}},
\end{gather}
where
\begin{gather}
C=\frac{8\,a}{3}e^3\left[e+(3e^2-1)\tanh^{-1}(e)\right]^{-1}.
\end{gather}
We need only determine the fluid stress in the resistance problem at the jet locations $\b{x}=\pm b\, \b{\hat{x}}$ in order to compute the velocity in the swimming problem. After some algebra, we find that (with $\b{\hat{n}}=\pm \b{\hat{x}}$ at the jet locations),
\begin{eqnarray}
&&(\b{\hat{n}}\cdot \bm{\tilde{\sigma}}\cdot\b{\hat{n}})(\b{x}=\pm b\,\b{\hat{x}})=-p+2\mu \left(\b{\hat{n}}\cdot \b{E}\cdot \b{\hat{n}}\right)=\\&&
\pm\frac{4\mu e \alpha}{a\sqrt{1-e^2}}\pm 2\mu\left(\frac{2 e \left(e^2-2\right) \left[e (e \tilde{U}-\alpha )+\left(\alpha -3 e^2 \alpha \right)\tanh^{-1}(e)\right]}{a \sqrt{1-e^2} \left[e+\left(3 e^2-1\right)\tanh^{-1}(e)\right]} \right)
\end{eqnarray}
Inserting the above into Eq.~\eqref{E:MainSwimmingVelocity} and simplifying, the swimming velocity of the jetting prolate ellipsoid is found to be 
\begin{gather}
U=-\frac{q}{2\pi a b}\cdot
\end{gather}

\subsection{Example 4: A viscous pinwheel}

A sufficiently asymmetric body can be made to rotate by appropriate placement of the jetting nozzles. The rotational velocity of a prolate spheroid driven about its minor axis, a ``viscous pinwheel,'' can be determined using Eq.~\eqref{E:MainSwimmingVelocity}. Consider the same prolate spheroid as described in the previous example, but with its major axis aligned with the $x$ axis. The body volume is $V(e)=\pi a^3 (1-e^2)$, with $e=\sqrt{a^2-b^2}/a$ the eccentricity. Pure rotation about the $z$ axis may be driven with outward flowing jets at $\b{x}_1=(\bar{x},-r(\bar{x}),0)$ and $\b{x}_2=(-\bar{x},r(\bar{x}),0)$, with $r(x)=\sqrt{1-e^2}\sqrt{a^2-x^2}$. In order to maintain internal volume conservation, suction jets (inward flowing jets) are placed symmetrically at $\b{x}_3=(\bar{x},r(\bar{x}),0)$ and $\b{x}_4=(-\bar{x},-r(\bar{x}),0)$, and we write $\phi(\b{x})=q(\delta(\b{x,x_1})+\delta(\b{x,x_2})-\delta(\b{x,x_3})-\delta(\b{x,x_4}))$. The flow at all four nozzles act with equal strength to drive the rotation. The corresponding unit normal vectors are $\b{\hat{n}_{o}}=(-r'(\pm\bar{x})\b{\hat{x}}\mp\b{\hat{y}})/\sqrt{1+r'(\bar{x})^2}$ at the outward flowing jets, and $\b{\hat{n}}_{i}=(-r'(\pm\bar{x})\b{\hat{x}}\pm\b{\hat{y}})/\sqrt{1+r'(\bar{x})^2}$ at the inward flowing jets.

\begin{figure}[htbp]
\begin{center}
\includegraphics[width=5.7in]{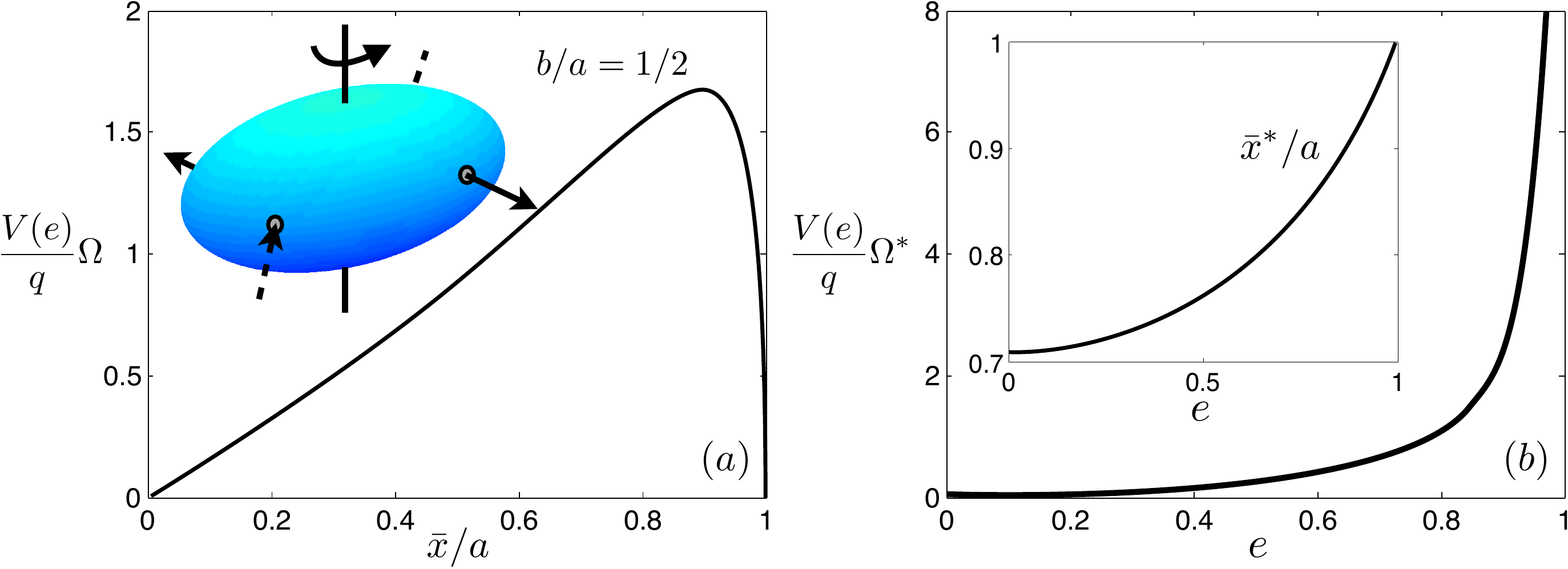}
\caption{(a) Rotational velocity response to two jet-pairs positioned symmetrically at $\pm \bar{x}$, for a body of aspect ratio $b/a=\sqrt{1-e^2}=1/2$. The rotational velocity is normalized by the relative flux per body volume. (b) Normalized maximal rotational velocity as a function of eccentricity $e$ (with $V(e)=\pi a^3 (1-e^2)$), and corresponding nozzle placement position (inset).}
\label{Figure3}
\end{center}
\end{figure}

The torque required to rotate the prolate spheroid about the $z$ axis with angular velocity $\tilde{\Omega}$ (the resistance problem) is
\begin{gather}
\b{\tilde{L}}=\frac{16\pi\mu (a\,e)^3(1-e^2)\tilde{\Omega}}{3\left[e-(1-e^2)\tanh^{-1}(e)\right]}\b{\hat{z}},
\end{gather}
(see Refs.~\cite{Jeffery22,cw74}). During this rotation, the pointwise fluid stress on the rotating solid, $\bm{\tilde{\sigma}}$, has an analytical expression which may be deduced using the singularity method of Chwang and Wu \cite{cw75} (though it is unwieldy and not included here). Assuming that the nozzles are small, only the normal component of the traction at the nozzle locations are relevant in setting the rotational velocity in the swimming/jetting problem, $\b{\Omega}=\Omega\,\b{\hat{z}}$, and so we have
\begin{gather}
\Omega=-\frac{3\left[e-(1-e^2)\tanh^{-1}(e)\right]}{16\pi\mu (a\,e)^3(1-e^2)\tilde{\Omega}}\sum_{i=1}^4 q(\b{\hat{n}}\cdot \bm{\tilde{\sigma}}\cdot \b{\hat{n}})(\b{x}_i).
\end{gather}
Figure~\ref{Figure3}a shows the rotational velocity of a spheroid with aspect ratio $b/a=1/2$ ($e\approx0.866$) as a function of the jet placement distance from the particle center along the major axis. Here $\Omega$ is normalized by the relative flux per body volume. The result confirms intuition: there is no rotation when the jets are placed symmetrically along either major or minor axis, and hence there exists an optimal jet placement for inducing a body rotation. Figure~\ref{Figure3}b shows the maximal angular speed, $\Omega^*$, which may be obtained by appropriate jet placement for bodies of eccentricity $e$, along with the optimal jet placement location $\bar{x}^*$ (inset). As the body becomes spherical the optimal jet placement limits to $\bar{x}^*/a=1/\sqrt{2}$, though the rotation response becomes increasingly small. As the body becomes more elongated the optimal jet placement moves closer to the poles, where the resulting forces can produce the largest torque on the body; along with the vanishing torque on a rotating slender body as $e\rightarrow 1$, the maximal rotational velocity increases without bound.

\subsection{Hydrodynamic efficiency}

We have shown that the velocities of a jetting body are dependent only upon the normal tractions in corresponding resistance problems, through Eq.~\eqref{E:MainSwimmingVelocity}. The issue of  hydrodynamic efficiency is rather more involved. We define here an efficiency used in many other works in low Reynolds number locomotion, and consider for simplicity only translational motion in what follows (see Ref.~\cite{Childress81}). The efficiency is defined as a ratio comparing the rate of mechanical work done onto the fluid in the resistance problem to that done in the swimming problem when the bodies are moving at the same velocity $\b{U}$,
\begin{gather}
\mathcal{E}=\frac{\b{U}\cdot \b{F}}{\displaystyle\int_{\partial D} \b{u}\cdot \left(\bm{-\sigma} \cdot \b{\hat{n}}\right)\,dS}\cdot\label{E:Efficiency}
\end{gather}
The difficulty in computing the efficiency, compared to simply computing the swimming velocity, is seen plainly in Eq.~\eqref{E:Efficiency}. Instead of integrating the fluid velocity in the swimming problem against the stress in the resistance problem, as in Eq.~\eqref{E:Lorentz}, here we must integrate against the stress in the swimming problem which depends intricately upon the precise form of the jetting flow profile. A theoretical bound on the above measure of $\mathcal{E}=75\%$ was found by Stone and  Samuel when the surface deformations act tangentially to the surface, and the body is spherical \cite{ss96}. Since $\mathcal{E}$ is a mechanical and not a thermodynamic efficiency, motion with $\mathcal{E}>1$ is theoretically possible, in particular when the no-slip condition does not hold everywhere upon the body surface and there are sources or sinks of fluid or surface material. For example, Leshansky et al. \cite{lk07} showed that a slender treadmilling spheroidal body can locomote with arbitrarily large efficiency by continuously introducing and removing surface material at the poles.

Since the body is force and torque free in the swimming problem, the efficiency still requires only knowledge of the force component normal to the porous surfaces, and we have
\begin{gather}
\mathcal{E}=\frac{\b{U}\cdot \b{F}}{\displaystyle \sum_{i=1}^m\int_{\partial D_\delta^i} \phi(\b{x})\left(-\b{\hat{n}}\cdot\bm{\sigma} \cdot \b{\hat{n}}\right)\,dS}\cdot\label{E:Efficiency2}
\end{gather}
As expected, the work done by the swimmer is the work done by the jets against the normal stresses in the fluid. Note that Eq.~\eqref{E:Efficiency2} is only the external  efficiency, and it does not include, for example, the internal work done to create the jetting flows in the first place. In the remainder of the paper we focus on axisymmetric spheroids, for which the swimming velocity and efficiency may be determined analytically, and we maximize the swimming efficiency through numerical optimization.

\section{Spherical Body Shape}

\subsection{General solution}

In the previous section we showed that the velocity of a spherical body is maximized by placing two small jets at the poles along the axis of locomotion. We now present an analysis of the dynamics and efficiency of a spherical body of radius $a$ with an arbitrary axisymmetric fluid-jetting profile. We consider only jetting profiles which are fore/aft asymmetric and therefore volume conserving. We also assume the jets, and the resulting flow, to be axisymmetric. Swimming spheres with arbitrary velocity boundary conditions have been studied in a more general setting by Lighthill \cite{Lighthill52} and Blake \cite{Blake71}. Magar and Pedley \cite{mp05}, and Ishikawa et al. \cite{isp06} have recently considered the behavior of swimming spheres with only tangential surface distortions, so-called {\em squirmers}, also using a similar analysis to that presented below.

\begin{figure}[b]
\begin{center}
\includegraphics[width=1.8in]{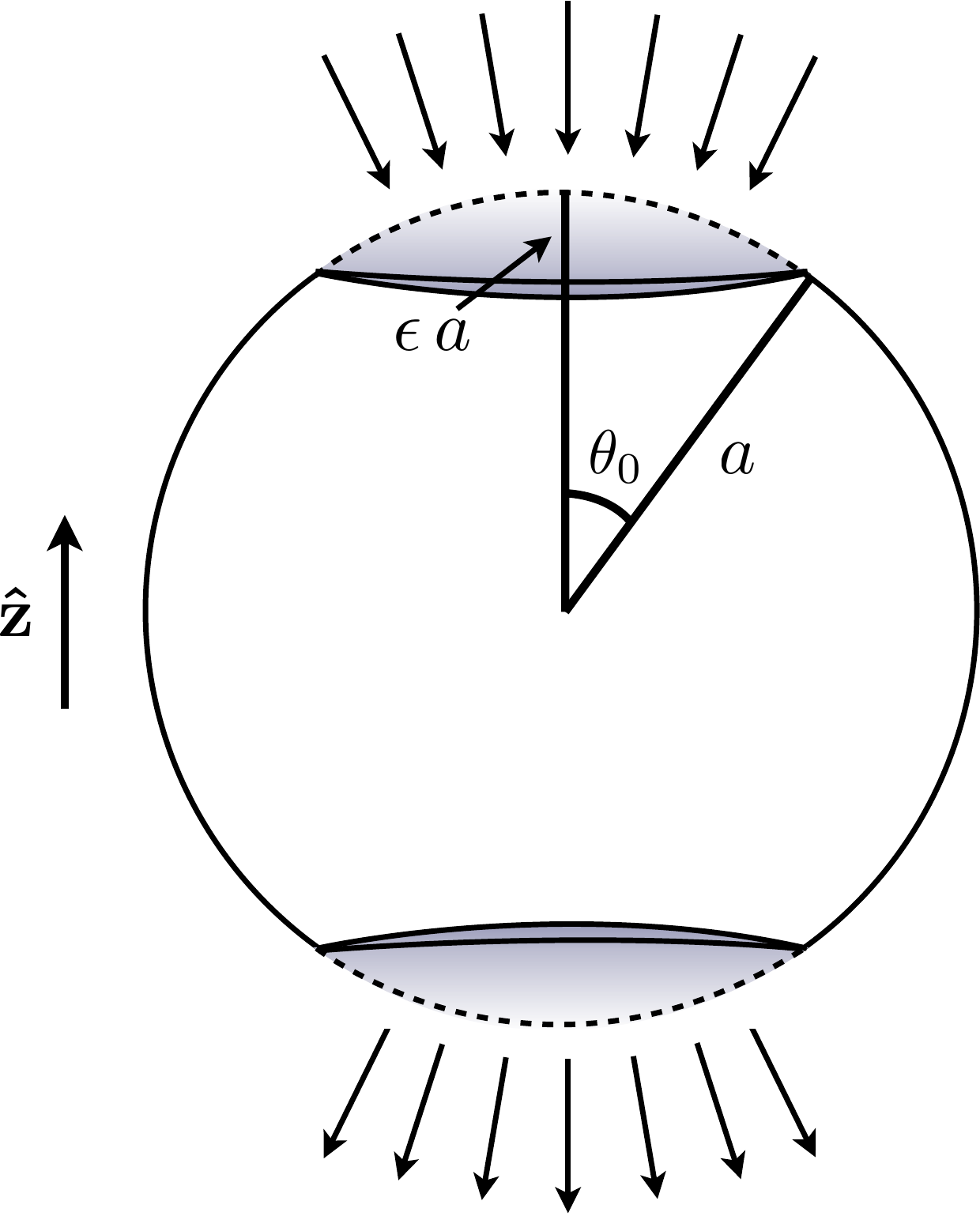}
\caption{A fluid-jetting sphere of radius $a$. The porous surfaces have equivalent spherical cap heights of lengths $\e\,a$, with $\e\in[0,1]$. The surface area of each porous cap is $S_\delta= 2\pi a^2 \e$, and $\zeta_0=\cos(\theta_0)=1-\e$. }
\label{Figure4}
\end{center}
\end{figure}

We set $\b{x}_1=\b{\hat{z}}$, $\b{x}_2=-\b{\hat{z}}$,  $S_\delta^1=S_\delta^2=S_\delta$, and $q=-q_1=q_2$ as before. Hence, the spherical body translates along the $\b{\hat{z}}$ direction and the Stokes equations may be solved in axisymmetric spherical coordinates $(r,\theta)$, with $r\in[a,\infty)$ the radial distance and $\theta\in[0,\pi]$ the polar angle. For notational convenience we again define $\zeta=\cos(\theta)$. The porous surfaces are taken to have spherical cap heights of lengths $\e\,a$, as illustrated in Fig.~\ref{Figure4}, with $\e\in[0,1]$. Hence, $S_\delta=2\pi a^2 \e $, and the polar angle illustrated in Fig.~\ref{Figure4} has $\zeta_0=\cos(\theta_0)=1-\e$. 

In order to determine the efficiency $\mathcal{E}$ we will solve the Stokes equations, Eqs.~\eqref{E:Stokes}-\eqref{E:Incompressibility}, with the general boundary conditions, Eqs.~\eqref{E:poreBC}-\eqref{E:BCs}. Exploiting the linearity of the Stokes equations, we decompose the problem into two parts, setting $\psi=\psi'+\psi_\delta$. The first problem (for $\psi'$) corresponds to axisymmetric rigid body motion in the swimming direction, while the second problem (for $\psi_\delta$) corresponds to the flow generated by a jetting sphere which is fixed in space centered at the origin. The separation of the problem into a rigid body motion and a jetting flow component is the same separation used to determine the swimming velocity in the previous section. 

The first problem has a well-known solution. The stream-function for flow due to a solid sphere translating through a quiescent fluid with velocity $U\b{\hat{z}}$ is
\begin{gather}
\psi'=\frac{U}{4}r^2\left(1-\zeta^2\right)\left[\left(\frac{a}{r}\right)^3-3\left(\frac{a}{r}\right) \right],
\end{gather}
with associated pressure and normal strain-rate tensor component (see \cite{hb65})
\begin{gather}
p'=\frac{3}{2a}\mu\,U\zeta,\,\,\,\,\b{\hat{r}}\cdot \b{E} \cdot \b{\hat{r}} = E'_{rr}=0.
\end{gather}

In order to solve the second problem for $\psi_\delta$, we write the fluid velocity in a fixed spherical coordinate system as $\b{u}=u_r\b{\hat{r}}+u_\theta\b{\hat{e}_\theta}$, and define an axisymmetric stream-function such that
\begin{gather}
u_r=-\frac{1}{r^2\sin(\theta)}\frac{\partial\psi_\delta}{\partial \theta},\,\,\,\,u_\theta=\frac{1}{r\sin(\theta)}\frac{\partial\psi_\delta}{\partial r}\cdot
\end{gather}
The use of the stream-function ensures that the incompressibility condition (Eq.~\ref{E:Incompressibility}) is automatically satisfied, and Eq.~\eqref{E:Stokes} becomes
\begin{gather}
D^2\left(D^2\psi_\delta\right)=0,\label{E:Stokes2}
\end{gather}
where
\begin{gather}
D^2=\frac{\partial^2}{\partial r^2}+\frac{1-\zeta^2}{r^2}\frac{\partial^2}{\partial \zeta^2}\cdot
\end{gather}
The boundary conditions on the body surface are the jetting profile in the direction normal to the surface, and zero tangential surface velocity,
\begin{gather}
u_r(r=a,\theta)=\phi(\zeta),\,\,\,\,u_\theta(r=a,\theta)=0.
\end{gather}
The general solution to Eq.~\eqref{E:Stokes2}, for flows which decay in the far-field and have bounded tangential velocities at the poles, may be written as
\begin{gather}
\psi_\delta=\displaystyle \sum_{n=2}^\infty \left[\frac{A_n}{r^{n-1}}+\frac{B_n}{r^{n-3}}  \right]\mathcal{G}_n(\zeta),
\end{gather}
with $\mathcal{G}_n(\zeta)$ the $n^{th}$ Gegenbauer function of the first kind \cite{Sampson91,hb65}. 

The Gegenbauer functions are related to the Legendre polynomials $P_n(\zeta)$ via
\begin{gather}
\mathcal{G}_n(\zeta)=\frac{P_{n-2}(\zeta)-P_n(\zeta)}{2n-1}
\end{gather}
for $n\geq 2$. Under this representation the radial and tangential components of velocity are written as
\begin{gather}
u_r=-\displaystyle \sum_{n=2}^\infty \left( \frac{A_n}{r^{n+1}}+\frac{B_n}{r^{n-1}}\right)P_{n-1}(\zeta),\label{E:ur}\\
u_\theta=-\displaystyle \sum_{n=2}^\infty \left( (n-1)\frac{A_n}{r^{n+1}}+(n-3)\frac{B_n}{r^{n-1}}\right)\frac{\mathcal{G}_{n}(\zeta)}{\sqrt{1-\zeta^2}}\cdot\label{E:ut}
\end{gather}
We denote the inner product of the radial velocity boundary condition on the body surface and the $n^{th}$ Legendre polynomial by $c_n$:
\begin{gather}
c_n=\langle u_r , P_n \rangle=\int_{-1}^1 \phi(\zeta) \,P_n(\zeta)\,d\zeta.
\end{gather}
Thus, by successive inner products of Eqs.~(\ref{E:ur}-\ref{E:ut}) we find
\begin{gather}
A_n=\frac{(2n-1)(n-3)}{4}a^{n+1}c_{n-1},\,\,\,\,\,\,B_n=-\frac{(2n-1)(n-1)}{4}a^{n-1}c_{n-1}.
\end{gather}
The corresponding pressure field, which may be found by integrating Eq.~\eqref{E:Stokes}, is
\begin{gather}
p=-\mu\displaystyle \sum_{n=2}^\infty\frac{2(2n-3)}{n}\frac{B_n}{a^n}P_{n-1}(\zeta),
\end{gather}
and the symmetric rate-of-strain tensor $\b{E}=\displaystyle\frac{1}{2}\left(\nabla \b{u}+\nabla \b{u}^T\right)$ has an $\b{\hat{r}\hat{r}}$ component
\begin{gather}
E_{rr}=\b{\hat{r}\cdot E \cdot \hat{r}}=\partial_r\,u_r \Big|_{r=a}=\displaystyle \sum_{n=2}^\infty \left((n+1) \frac{A_n}{a^{n+2}}+(n-1)\frac{B_n}{a^{n}}\right)P_{n-1}(\zeta).
\end{gather}

\subsection{Efficiency and optimization of an entirely porous sphere}
The swimming efficiency $\mathcal{E}$ from Eq.~\eqref{E:Efficiency} may now be determined by combining the two problems, $\psi=\psi'+\psi_\delta$. The rate of mechanical work performed on the fluid in the full swimming problem is
\begin{gather}
\Phi=\int_{\partial D} \b{u} \cdot(-\bm{\sigma} \cdot \b{\hat{n}})\,dS=-2\pi a^2\int_{-1}^1 \phi(\zeta) \left[-(p+p')+2\mu E_{rr} \right]\,d\zeta,
\end{gather}
where we have used that the swimming body imparts no force or torque onto the fluid. Inserting the expressions from above, we have
\begin{gather}
-(p+p')+2\mu E_{rr}=\mu  \displaystyle \sum_{n=2}^\infty \left[2(n+1) \frac{A_n}{a^{n+2}}+\left(2n+2-\frac{6}{n}\right)\frac{B_n}{a^n} \right]P_{n-1}(\zeta)-\frac{3}{2a}\mu\,U\,\zeta,
\end{gather}
and after further simplification and a shift in the summation we find
\begin{gather}
\Phi=\displaystyle \pi\,a\,\mu \displaystyle \sum_{n=1}^\infty (2n+1)\left[(2n+1)+\frac{3}{n+1} \right]c_{n}^2+3\pi\mu \,a\,U\int_{-1}^1\zeta\, \phi(\zeta)\,d\zeta.
\end{gather}
The swimming velocity, from Eq.~\eqref{E:MainSwimmingVelocity}, and related towing force are
\begin{gather}
U=-\frac{1}{4\pi a^2}\int_{\partial D}\phi(\b{x})\left(\b{\hat{z}\cdot \hat{n}(\b{x})}\right)\,dS=-\frac{1}{2}\int_{-1}^1\zeta\,\phi(\zeta)\,d\zeta,\\
F=6\pi\mu \,a\, U=-3\pi\mu\,a\int_{-1}^1\zeta\,\phi(\zeta)\,d\zeta.
\end{gather}
The sphere moves with a non-zero velocity if and only if the jetting profile $\phi(\zeta)$ contains the first Legendre polynomial mode. Moreover, this is the only component of the jetting velocity that contributes to the swimming speed. The hydrodynamic efficiency, on the other hand, does depend upon the full nature of the jetting profile $\phi(\zeta)$, and we find
\begin{gather}
\mathcal{E}=\frac{\displaystyle \left(\int_{-1}^1\,\zeta \,\phi(\zeta)\,d\zeta\right)^2}{8\left(\displaystyle \int_{-1}^1\zeta\, \phi(\zeta)\,d\zeta\right)^2+\displaystyle \frac{2}{3} \displaystyle \sum_{n=2}^\infty (2n+1)\left(2n+1+\frac{3}{n+1} \right)\left(\int_{-1}^1P_n(\zeta)\phi(\zeta)\,d\zeta\right)^2}\cdot\label{E:SphereE}
\end{gather}
As a simple example consider a sphere which is entirely porous ($\e=1$), which is propelled with the jetting profile
\begin{gather}\label{linear}
\phi(\zeta)=\frac{q}{\pi a^2}P_1(\zeta)=\frac{q}{\pi a^2}\zeta.
\end{gather}
The jetting profile satisfies the flux constraints
\begin{gather}
\int_{\partial D}\phi(\b{x})\,dS=2\pi a^2\int_{-1}^1\phi(\zeta)\,d\zeta=0,\,\,\,\,\,\,\,\,\int_{\partial D_\delta^1}\phi(\b{x})\,dS=q,\,\,\,\,\,\,\,\,\int_{\partial D_\delta^2}\phi(\b{x})\,dS=-q.\label{E:Fluxconstraints}
\end{gather}

\begin{figure}[t]
\begin{center}
\includegraphics[width=4in]{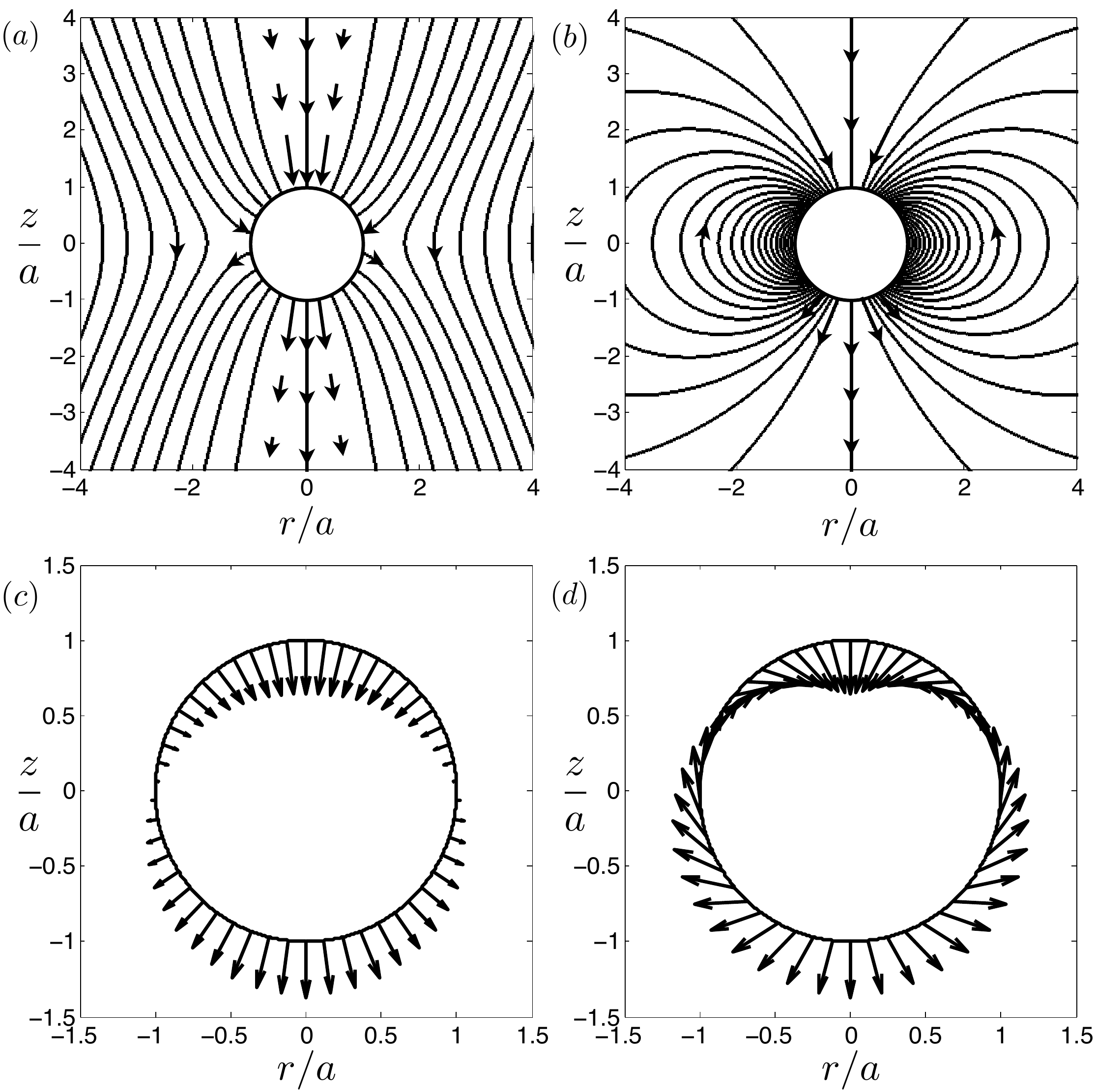}
\caption{The optimal jetting sphere (cross-section), with $\phi(\zeta)=P_1(\zeta)/S_\delta$. (a) Streamlines for a jetting sphere which is fixed at the origin, $\b{u}(r=a,\theta)=\phi(\zeta)\b{\hat{n}}$. (b) Streamlines for a swimming sphere, $\b{u}(r=a,\theta)=U\b{\hat{z}}+\phi(\zeta)\b{\hat{n}}$. (c) Distribution of surface fluid velocity for the fixed body, as in (a). (d) Distribution of surface velocity for the swimming body, as in (b).}
\label{Figure5}
\end{center}
\end{figure}

Streamlines generated by the jetting body held fixed at the origin are shown in Fig.~\ref{Figure5}a (for $q<0$), and in Fig.~\ref{Figure5}b when the body is swimming freely through the fluid. The dipolar fluid structure in the latter is evident. The distribution of fluid velocities at the body surface in each case are shown in Figs.~\ref{Figure5}c-d. The corresponding swimming velocity and hydrodynamic efficiency are
\begin{gather}
U=-\frac{q}{3\pi a^2},\,\,\,\,\mathcal{E}=\frac{1}{8}=12.5\%\,\cdot
\end{gather}

\begin{figure}[t]
\begin{center}
\includegraphics[width=2.0in]{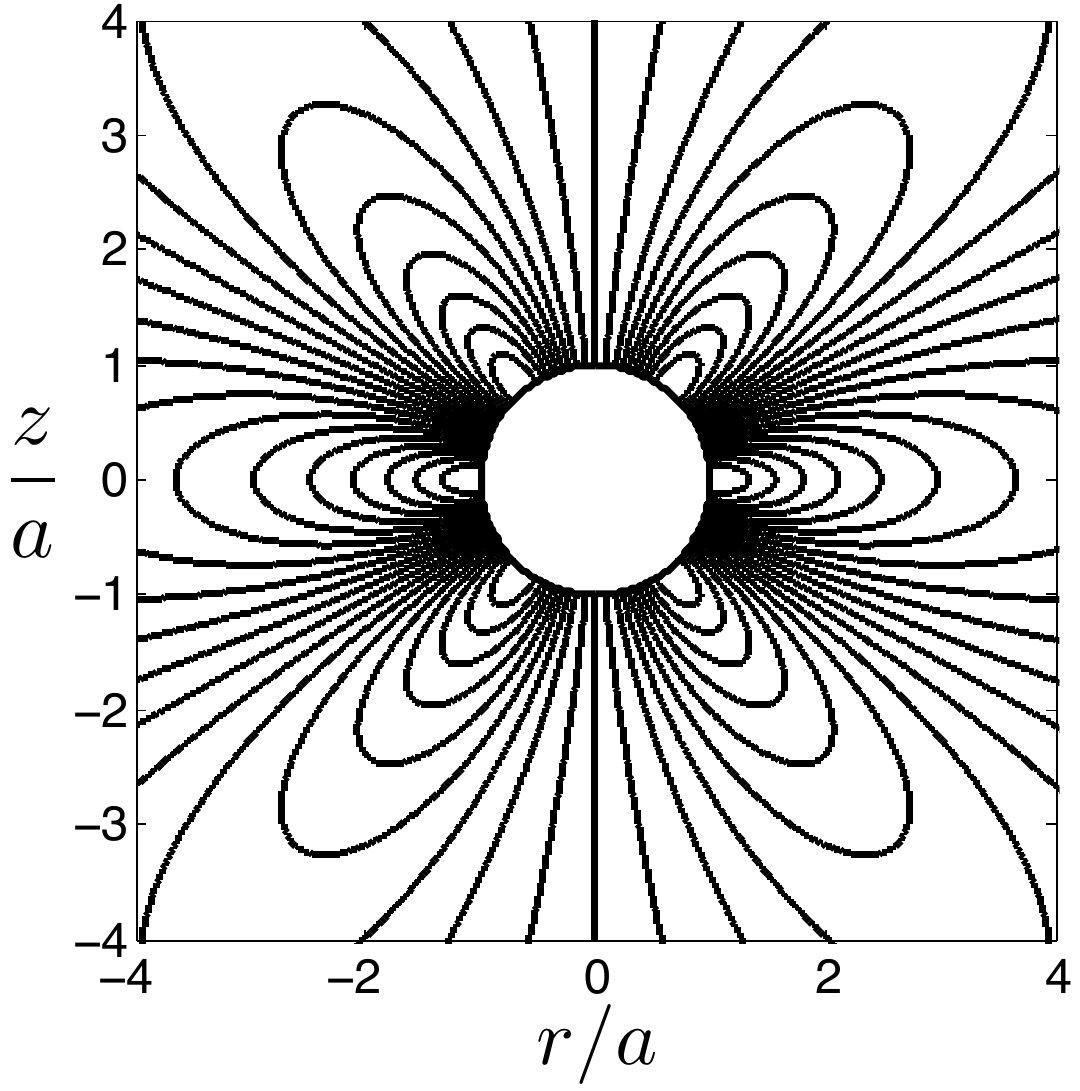}
\caption{Streamlines for flow generated by jetting profile $\phi(\zeta)\propto P_3(\zeta)$.}
\label{Figure6}
\end{center}
\end{figure}

The swimming speed is smaller than the largest possible swimming speed as discussed in the previous section by a factor of $2/3$. However, it is simple to see that this jetting profile is in fact the most hydrodyamically efficient jetting profile possible. Every component of $\phi(\zeta)$ which has a non-zero inner product with the $n^{th}$ Legendre polynomial, for $n\geq 2$, contributes positively to the denominator of $\mathcal{E}$ in Eq.~\eqref{E:SphereE}, without increasing the swimming velocity. Hence, the optimal surface jet pattern will only have the first Legendre polynomial component. Moreover, the decrease in the efficiency with the inclusion of any other Legendre modes in the jetting profile is dramatic, as can be seen by the form of Eq.~\eqref{E:SphereE}. The higher Legendre modes do not contribute to the swimming velocity, but introduce large wave-number variations in the fluid flow, corresponding to an increased viscous dissipation of energy into the fluid. As an example, streamlines associated with the third Legendre mode are shown in Fig.~\ref{Figure6}, from which the corresponding increase in viscous dissipation may be intuited.

\subsection{Efficiency and optimization of a partially porous sphere}

Organisms or manmade swimming devices utilizing non-inertial jet propulsion might be constrained to draw in and expel fluid from only some parts of the surface. Here we consider the efficiency and optimal jetting profiles of spherical bodies with $\e < 1$ (see Fig.~\ref{Figure4}).

The efficiency derived in the previous section, Eq.~\eqref{E:SphereE}, is general and applies to such a partially porous body. We first note that the swimming efficiency is zero if the surface fluid velocity is discontinuous in $\zeta$. Physically, this corresponds to solutions with non-integrable stress singularities at the edge of the pore (the Legendre modes needed to represent $\phi(\zeta)$ do not decay at a sufficient rate to give a finite value of the efficiency). A discontinuous jetting profile is considered in Appendix A. The Legendre modes decay as $c_n\sim n^{-3/2}$, and as expected the sum in the denominator of Eq.~\eqref{E:SphereE} diverges. 
\begin{figure}[t]
\begin{center}
\includegraphics[width=4.5in]{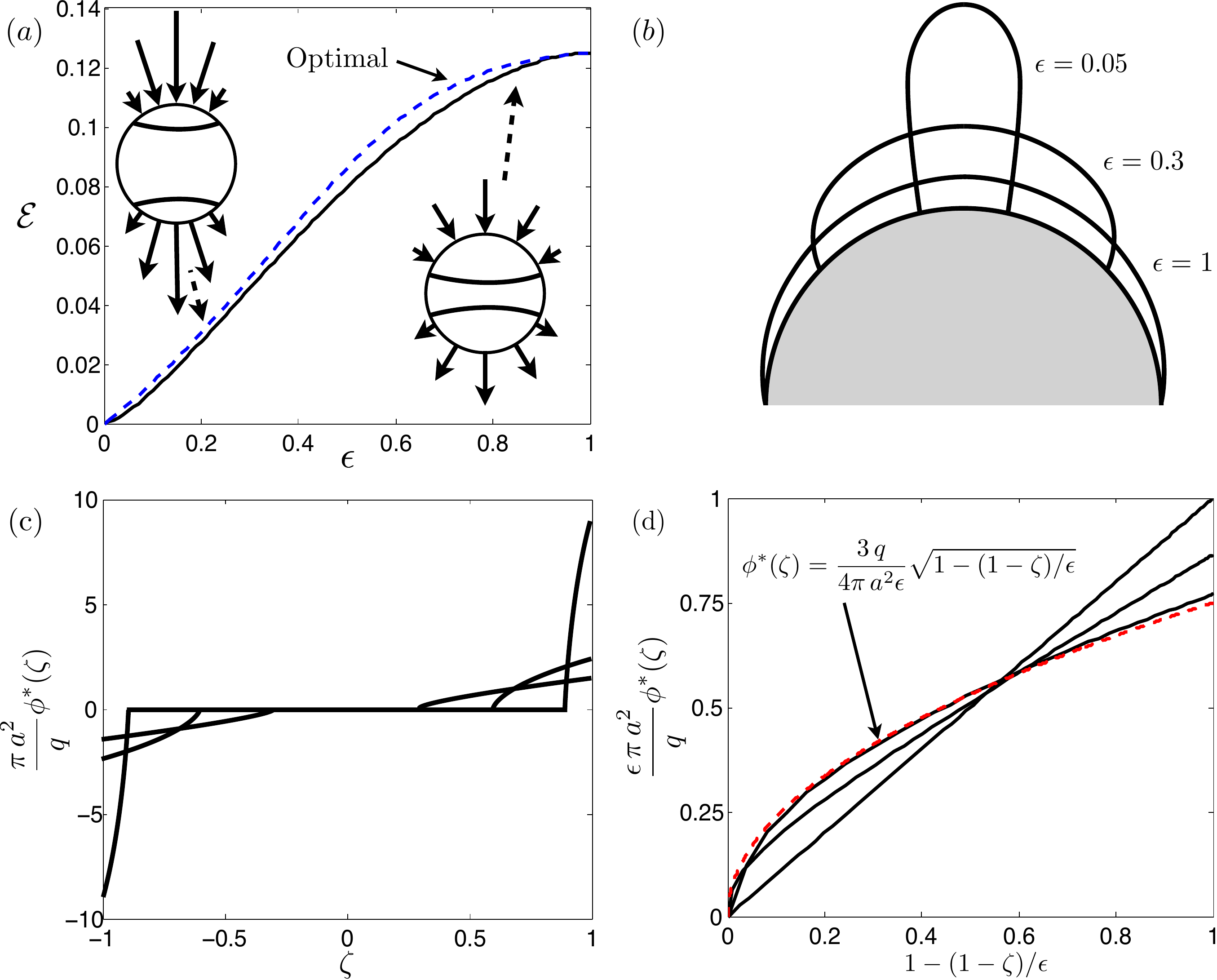}
\caption{(a) Efficiency as a function of $\e$ for the jetting profile $\phi(\zeta)=\displaystyle\frac{q}{\pi a^2 \e}\Big[\left(1-(1-\zeta)/\e\right)\chi_{\{\zeta\in[1-\e,1]\}}-(1-(1+\zeta)/\e)\chi_{\{\zeta\in[-1,-1+\e]\}}\Big]$ as a solid line, and the numerically determined optimal efficiency as a dashed line. (b) Optimal jetting profiles. Relative distance from the body indicates jetting flow velocity at the surface. Individual profiles are scaled for presentation purposes. (c) Optimal jetting profiles, $\phi^*(\zeta)$ as a function of $\zeta$ for $\e=0.5$, 
$\e=0.2$, and $\e=0.1$. (d) Same as in (c), but scaled. The optimal flow profile through an increasingly small porous surface area appears to limit to the noted function and is shown as a dashed line.}
\label{Figure7}
\end{center}
\end{figure}

To see the general scaling of the efficiency with the porous surface area, consider the following form for the jetting profile which has $\phi(\zeta)=0$ at the porous surface edges (so that the surface fluid velocity is continuous on $\zeta=\cos(\theta)\in[-1,1]$):
\begin{gather}
\phi(\zeta)=\frac{q}{\pi a^2\,\e}\left[\left(1-\frac{(1-\zeta)}{\e}\right)\chi_{\{\zeta\in[1-\e,1]\}}-\left(1-\frac{(1+\zeta)}{\e}\right)\chi_{\{\zeta\in[-1,-1+\e]\}}\right],
\end{gather}
where $\chi$ is the indicator function, which would correspond to the quadratic profile of pressure driven flow through a single open pore as $\e\rightarrow 0$ (see Ref.~\cite{hb65}). The corresponding swimming velocity is
\begin{gather}
U=-\frac{1}{2}\int_{-1}^1\zeta\phi(\zeta)\,d\zeta=\frac{-(3-\epsilon )q}{6\pi a^2}, 
\end{gather}
and we have
\begin{gather}
c_n=\frac{q(1-(-1)^n)}{\pi a^2\,\e}\int_{1-\e}^1\left(1-\frac{(1-\zeta)}{\e}\right)\,P_n(\zeta)\,d\zeta.
\end{gather}

The resulting efficiency $\mathcal{E}$ as a function of $\e$ is shown in Fig.~\ref{Figure7}a as a solid line. The efficiency is monotonically increasing with the size of the porous cap height, and limits to the optimal value for a spherical jetting body of $\mathcal{E}=1/8=12.5\%$ as the body becomes entirely porous ($\e\rightarrow 1$). Even for $\e=0.2$ ($\theta_0\approx 36^\circ$) the body achieves efficiencies on the same order as those mechanisms employed by biological organisms, $\mathcal{E}\approx 3\%$.

The optimal jetting profile cannot be determined as directly as in the entirely porous case, which was determined by simple inspection of Eq.~\eqref{E:SphereE}. Instead, we seek numerically the optimal jetting profile by discretizing $\phi(\zeta)$ at points corresponding to the Gaussian quadrature nodes of order $M$, and computing the efficiency via Eq.~\eqref{E:SphereE} keeping the first $N$ terms in the summation. The profile is subject to the constraints of Eq.~\eqref{E:Fluxconstraints}, and $\phi(\zeta)=0$ for $|\zeta| \leq 1-\e$. The optimal profile is selected using an SQP quasi-Newton line search method built into the MATLAB optimization toolbox. The number of discritization nodes $M$ and summation terms $N$ are increased until there is no discernible variation in the result. 

The optimal jetting profiles so obtained, denoted by $\phi^*(\zeta)$, are shown in Figs.~\ref{Figure7}b-d, and correspond to the efficiencies shown as a dashed line in Fig.~\ref{Figure7}a. The entirely porous case, $\e=1$, returns the expected result, a profile which scales linearly with $\zeta$. Under the scaling indicated by the axes in Fig.~\ref{Figure7}d, it appears that the optimal jetting distribution through small pores at the poles ($\e\rightarrow 0$) is limiting to the form $\phi(\zeta)=3q/(4\pi a^2 \e)\sqrt{1-(1-\zeta)/\e}$, which is shown as a dashed line. The maximal efficiency shown in Fig.~\ref{Figure7}a does not deviate dramatically from the example of previous consideration. 

\section{Prolate Body Shape}

The efficiency of swimming at low Reynolds number is known to depend significantly upon body shape. The drag on a solid body of a given volume is minimized when the shape is approximately a prolate spheroid with aspect ratio $\approx 1/2$, but with conical endpoints \cite{Pironneau73,Bourot74}. Swimming organisms can enjoy a decreased fluid drag by selecting a similarly streamlined shape. However, in a departure from the more common shape-drag relationship, the propulsive mechanism of a fluid-jetting body is such that the efficiency decreases as the body becomes more prolate, as we shall show. We begin by solving for the efficiency exactly for one particular jetting profile, then move on to consider a more general framework for studying the fluid-jetting dynamics and efficiency. Finally, as in the case of a spherical body shape, we will determine numerically the optimal jetting profile for both entirely and partially porous jetting surfaces. 

\subsection{An exact solution}

We proceed just as in the spherical case, but the Stokes equations are now solved in prolate spheroidal coordinates ~\cite{hb65}. The appropriate coordinate system is given by the conformal mapping
\begin{gather}
z+i\,r=c\cosh(\xi+i\,\theta),
\end{gather}
and we set
\begin{gather}
\tau=\cosh(\xi),\,\,\,\,\,\zeta=\cos(\theta)
\end{gather}
for clarity. The surface defined by $\tau=\tau_0>1$ is a confocal spheroid with foci $\pm c\,\b{\hat{z}}$, and we have $\tau\in[1,\infty)$, $\zeta\in[-1,1]$. For a spheroid with major and minor axis lengths $2a$ and $2b$, respectively, we have $\tau_0=a/c=a/\sqrt{a^2-b^2}$, the inverse eccentricity as in \S II. The Stokes stream-function satisfies the biharmonic equation
\begin{gather}
D^4 \psi=0,\label{E:StokesStream}
\end{gather}
where
\begin{gather}
D^2=\frac{1}{c^2\left(\tau^2-\zeta^2\right)}\Big[\left(\tau^2-1\right)\partial_{\tau\tau}+\left(1-\zeta^2\right)\partial_{\zeta\zeta} \Big].
\end{gather}
Writing the velocity in the new coordinates, $\b{u}=u_\tau \bm{\hat{\tau}}+ u_\zeta \bm{\hat{\zeta}}$, where
\begin{gather}
\bm{\hat{\tau}}=\frac{\tau\sqrt{1-\zeta^2}}{\sqrt{\tau^2-\zeta^2}}\b{\hat{r}}+\frac{\zeta\sqrt{\tau^2-1}}{\sqrt{\tau^2-\zeta^2}}\b{\hat{z}},\\
\bm{\hat{\zeta}}=-\frac{\zeta\sqrt{\tau^2-1}}{\sqrt{\tau^2-\zeta^2}}\b{\hat{r}}+\frac{\tau\sqrt{1-\zeta^2}}{\sqrt{\tau^2-\zeta^2}}\b{\hat{z}},
\end{gather}
we set
\begin{gather}
u_\tau=\frac{1}{c^2\sqrt{\left(\tau^2-\zeta^2\right)\left(\tau^2-1\right)}}\frac{\partial \psi}{\partial \zeta},\,\,\,\,u_\zeta=-\frac{1}{c^2\sqrt{\left(\tau^2-\zeta^2\right)\left(1-\zeta^2\right)}}\frac{\partial \psi}{\partial \tau}\label{E:uProlate},
\end{gather}
so that the incompressibility condition is automatically satisfied. Also of use are the relations $\partial_{\xi}=\left(\tau^2-1\right)^{1/2}\partial_\tau$ and $\partial_{\eta}=-\left(1-\zeta^2\right)^{1/2}\partial_\zeta$. The surface area element is $dS=J(\zeta)\,d\zeta\,d\theta$, with $J(\zeta)=c^2\sqrt{(\tau_0^2-1)(\tau_0^2-\zeta^2)}=b\sqrt{a^2-(a^2-b^2)\zeta^2}$. 

An exact solution can be derived for the particular jetting profile
\begin{gather}
\phi(\zeta)=\frac{q}{\pi}\frac{\zeta}{J(\zeta)}=\frac{q}{\pi}\frac{\zeta}{c^2\sqrt{(\tau_0^2-1)(\tau_0^2-\zeta^2)}},\label{E:prolateexample}
\end{gather}
which limits to the linear profile as studied in the spherical case, Eq.~\eqref{linear},  as $b\rightarrow a$. 
Once again we split the problem into two parts, $\psi=\psi'+\psi_\delta$. The first problem for $\psi'$ is again that of rigid body motion along the axis of symmetry with swimming velocity $U\b{\hat{z}}$. Here also the solution is known \cite{hb65}, and we have
\begin{gather}
 \psi'=-\frac{U\,c^2}{2}(1-\zeta^2)(\tau^2-1)\frac{\left(\tau_0^2+1\right)/\left(\tau_0^2-1\right)\log\left[\displaystyle\frac{\tau +1}{\tau -1}\right]-\displaystyle\frac{2\tau }{\tau ^2-1}}{\left(\tau_0^2+1\right)/\left(\tau_0^2-1\right)\log\left[\displaystyle\frac{\tau_0+1}{\tau_0-1}\right]-\displaystyle\frac{2\tau_0}{\tau_0^2-1}}\cdot\label{E:psiprimepro}
\end{gather}
The corresponding pressure field may be found by integrating Eq.~\eqref{E:Stokes} in the appropriate coordinates, leading to
\begin{gather}
\frac{\partial p}{\partial \tau}=\frac{\mu}{c(\tau^2-1)}\partial _{\zeta}\left(D^2 \psi\right),\label{E:Pressures1}\\
\frac{\partial p}{\partial \zeta}=-\frac{\mu}{c(1-\zeta^2)}\partial _{\tau}\left(D^2 \psi\right),\label{E:Pressures2}
\end{gather}
yielding, upon integration,
\begin{gather}
p'(\tau,\zeta)=\frac{2 \,\mu\,U \zeta }{c \left(\tau^2-\zeta^2\right) \left[\left(\tau_0^2+1\right)\coth^{-1}(\tau_0)- \tau_0\right]}\cdot\label{E:pPrime}
\end{gather}
The $\bm{\hat{\tau}\hat{\tau}}$ component of the symmetric rate-of-strain tensor is
\begin{gather}
E'_{\tau\tau}=\bm{\hat{\tau}}\cdot \b{E}'\cdot \bm{\hat{\tau}}=\frac{\sqrt{\tau^2-1}}{c(\tau^2-\zeta^2)^{1/2}}\frac{\partial u_\tau}{\partial \tau}-\frac{\zeta\sqrt{1-\zeta^2}\,u_\zeta }{c \left(\tau ^2-\zeta ^2\right)^{3/2}}\cdot\label{E:Etautau}
\end{gather}
Inserting Eq.~\eqref{E:uProlate} and simplifying, we find on the surface $\tau=\tau_0$ that $E'_{\tau\tau}(\tau_0,\zeta)=0$. The pressure above corresponds to the stress used in the example of a prolate spheroid translating along its axis of symmetry in \S II. As shown in that example, the stress on the body during rigid body motion is sufficient information for determining the swimming speed in the full swimming problem. 

From Eq.~\eqref{E:MainSwimmingVelocity}, we have as in the example of \S II (replacing $U$ with $\tilde{U}$ in $p'$ and $E'_{\tau\tau}$),
\begin{gather}
\tilde{U} U=\frac{-1}{6\pi \mu R}\int_{\partial D} \phi(\b{x})\left(-p'+2\mu E'_{\tau \tau} \right)\,dS=\frac{-c^2}{3 \mu R}\int_{-1}^1 \phi(\b{\zeta})\left(-p'\right)\Big|_{\tau=\tau_0}\sqrt{(\tau_0^2-1)(\tau_0^2-\zeta^2)}\,d\zeta,
\end{gather}
leading to the general expression
\begin{gather}
U=-\frac{1}{2}\int_{-1}^1 G(\zeta)\phi(\b{\zeta})\,d\zeta\label{E:UProlateG},\quad 
G(\zeta)=\zeta\,\sqrt{\frac{\tau_0^2-1}{\tau_0^2-\zeta^2}}.
\end{gather}
As $b\rightarrow a$, we recover $G(\zeta)=\zeta$ as found for the spherical body. Inserting the jetting profile from Eq.~\eqref{E:prolateexample},
we finally obtain \begin{gather}
U=-\frac{q \,\tau_0^2[\tau_0\coth^{-1}(\tau_0)-1]}{a^2 \pi}\cdot\label{E:Upro}
\end{gather}
The velocity limits to that of the spherical body as $\tau_0\rightarrow \infty$, and increases without bound (for fixed $q$) as the body becomes infinitely slender. 

In order to determine the efficiency, we must now solve the second problem (for $\psi_\delta$), that of a jetting spheroid fixed in space at the origin,
\begin{gather}
D^4 \psi_\delta = 0, \label{E:pde}\\
\partial_\zeta \psi_\delta(\tau_0,\zeta)=J(\zeta)\phi(\zeta),\,\,\,\,\partial_\tau \psi_\delta(\tau_0,\zeta)=0,\,\,\,\,\,\psi_\delta(\tau\rightarrow \infty,\zeta)=0.\label{E:pde2}
\end{gather}
The form of the velocities on the body surface suggest the following ansatz for the stream function,
\begin{gather}
\psi_\delta=(1-\zeta^2)g(\tau),
\end{gather}
and after some manipulations we arrive at the solution,
\begin{gather}
\psi_\delta=\frac{q}{2\pi} \left(1-\zeta ^2\right)\frac{ \left[\tau +\left(\tau ^2-1\right) \coth^{-1}(\tau)-2 \tau  \tau_0\coth^{-1}(\tau_0)\right]}{\left(\tau_0^2+1\right)\coth^{-1}(\tau_0)-\tau_0}\cdot
\end{gather}
The corresponding surface pressure is, upon integration of Eqs.~\eqref{E:Pressures1}-\eqref{E:Pressures2},
\begin{gather}
p(\tau_0,\zeta)=\frac{2\, q\, \zeta\, \mu\,[\tau_0\coth^{-1}(\tau_0)-1]}{c^3 \pi  \left(\tau_0^2-\zeta ^2\right) \left[\left(\tau_0^2+1\right)\coth^{-1}(\tau_0)-\tau_0\right]},\label{E:pp2}
\end{gather}
and the normal component of the traction is (from Eq.~\eqref{E:Etautau}),
\begin{gather}
E_{\tau\tau}=\frac{q\, \zeta\,  \tau_0\left(1+\zeta^2-2\tau_0^2\right)}{c^3 \pi  \left(\tau_0^2-\zeta^2\right)^2 \left(\tau_0^2-1\right)}\cdot
\end{gather}

We now have the required information for determining the hydrodynamic efficiency in the full swimming problem. The work done on the fluid by the swimming body is
\begin{eqnarray}
\Phi&=&-2\pi\int_{-1}^1\phi(\zeta)(-p+2\mu E_{\tau\tau}-p')J(\zeta)\,d\zeta\\
&=&-\frac{4\mu q^2}{c^3\pi}\int_{-1}^1\frac{\zeta  \tau_0\left(1+\zeta^2-2 \tau_0^2\right)}{\left(\tau_0^2-\zeta ^2\right)^2 \left(\tau_0^2-1\right)}\zeta\,d\zeta=\frac{4\mu q^2}{c^3\pi}\left[\frac{\tau_0-\left(\tau_0^2+1\right)\coth^{-1}(\tau_0)}{\tau_0^2-1}\right].
\end{eqnarray}
Meanwhile, using $\b{\tilde{F}}=6\pi\mu R \b{\tilde{U}}$ with $R$ as defined in Eq.~\eqref{E:Rdef}, we find the efficiency
\begin{gather}
\mathcal{E}=\frac{2 \left(\tau_0^2-1\right) (\tau_0\coth^{-1}(\tau_0)-1)^2}{\left[\tau_0-\left(\tau_0^2+1\right) \coth^{-1}(\tau_0)\right]^2}\cdot\label{E:Epro}
\end{gather}

The efficiency is found to decrease monotonically to zero as the body becomes more slender, as shown in Fig.~\ref{Figure9}a as a dashed line. In the spherical limit (as $b\rightarrow a$) we recover $\mathcal{E}\rightarrow 12.5\%$ as expected. 

As previously noted, it is unusual that a moving body's hydrodynamic efficiency decreases monotonically with its slenderness. The reader may have already guessed this result, however, given the nature of the propulsive mechanism. The jetting flow through the body is constrained to act locally in the direction normal to the body surface. As the body becomes more prolate in shape, the nozzles on the surface become oriented in a direction more perpendicular to the direction of motion, and become much less useful for propulsion. From this observation, it is thus expected that the efficiency will increase as the body becomes more oblate in shape, even though the surface area presented to the fluid is increasing in that case. This will be the subject under consideration in \S V. First, though, we develop a general framework for studying prolate bodies with arbitrary jetting profiles, and determine numerically the optimal actuation for both entirely and partially porous body surfaces.

\subsection{Efficiency and optimization of an entirely porous body}

The dynamics and efficiency of a prolate jetting body for an arbitrary jetting profile, and for partially porous surfaces, requires a different approach to solving the jetting problem for $\psi_\delta$. The general solution to Eqs.~\eqref{E:pde}-\eqref{E:pde2} may be expressed using spheroidal harmonic functions $R_n$ and $S_n$, themselves expressed in terms of the Legendre polynomials of the first and second kind, $P_n$ and $Q_n$ respectively (see \cite{kst09}):
\begin{gather}
\psi_\delta=A_1\,R_1(\tau)R_1(\zeta)+\displaystyle\sum_{n\geq 1}\left[B_n+\tilde{B}_n\left(\tau^2+\zeta^2\right)\right]R_n(\zeta)S_n(\tau),\label{E:PsiProlate}
\end{gather} 
where
\begin{eqnarray}
R_n(x)&=&(1-x^2)P_n'(x)=-n\,x P_n(x)+n\,P_{n-1}(x),\label{E:Rn}\\
S_n(x)&=&(1-x^2)Q_n'(x)=-n\,x Q_n(x)+n\,Q_{n-1}(x),\\
Q_n(x)&=&P_n(x)\int_{x}^\infty\frac{d\lambda}{P_n(\lambda)^2(\lambda^2-1)}\cdot
\end{eqnarray}

As discussed in more detail in Appendix B, the boundary conditions are used to determine the coefficients $B_n$ and $\tilde{B}_n$. Subsequently, the pressure is found through integration of Eqs.~\eqref{E:Pressures1}-\eqref{E:Pressures2}, and for arbitrary $\phi(\zeta)$ we find that
\begin{gather}
E_{\tau\tau}(\tau_0,\zeta)=\left[\frac{\tau_0(1+\zeta^2-2\tau_0^2)}{c(\tau_0^2-\zeta^2)^{3/2}(\tau_0^2-1)^{1/2}}\right]\phi(\zeta).
\end{gather}

Meanwhile, the solution to the rigid body problem for $\psi'$ (Eq.~\ref{E:psiprimepro}) does not depend on the jetting profile outside of the dependence of $U$ on $\phi(\zeta)$, and the results of the previous section carry over here. For a given jetting profile $\phi(\zeta)$, the first $N$  coefficients $B_n$ and $\tilde{B}_n$ are determined numerically and the efficiency is then found using a Gaussian quadrature (see Appendix B).

\begin{figure}[t]
\begin{center}
\includegraphics[width=5.0in]{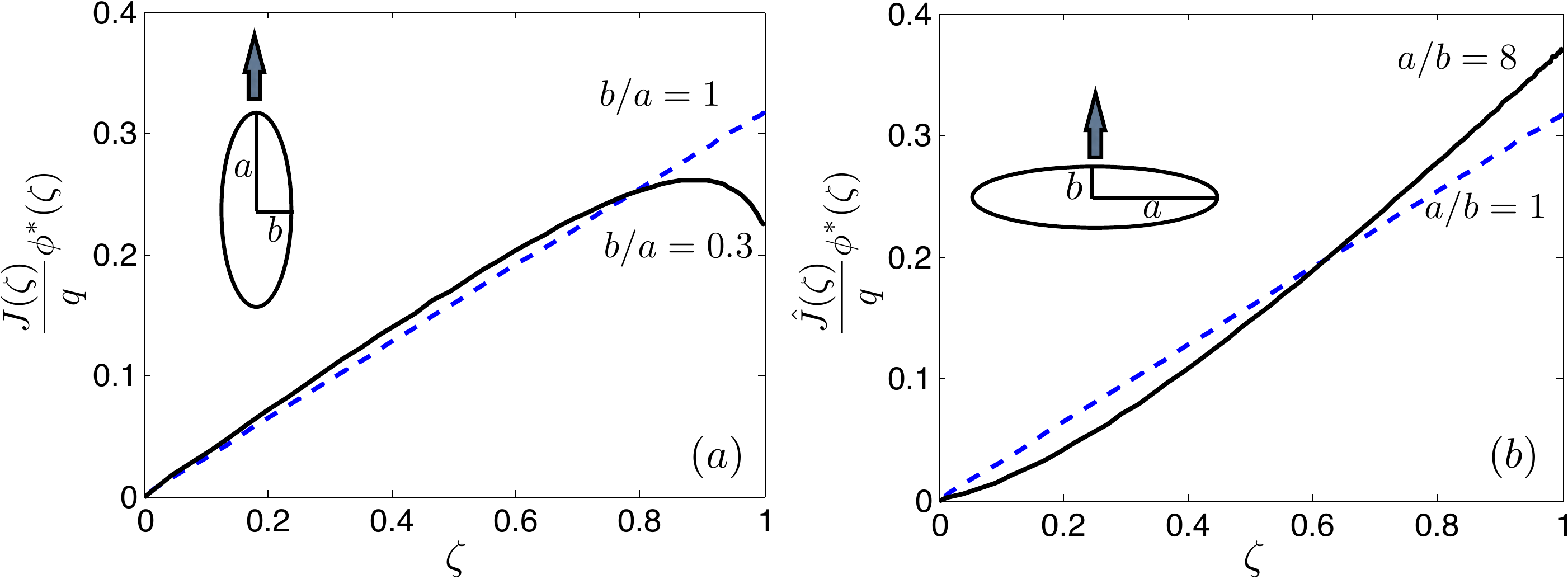}
\caption{(a) The optimal geometry-normalized jetting profiles for a sphere (dashed line) and a prolate spheroidal body (solid line). (b) The same, for a sphere (dashed line) and an oblate spheroidal body (solid line).}
\label{Figure8}
\end{center}
\end{figure}

Given the framework described above, the optimal jetting profile was selected by a similar SQP quasi-Newton line search method as the one used for optimizing the spherical swimmer. The optimal profile $\phi^*(\zeta)$ so found (normalized by the geometric scaling $J(\zeta)/q$) is shown for a body of aspect ratio $b/a=0.3$ in Fig.~\ref{Figure8}a, along with the optimal profile for the spherical case ($b/a=1$). As the body becomes more slender in shape, the only jets which contribute appreciably to the swimming velocity, via Eq.~\eqref{E:UProlateG}, are those near the fore and aft poles. However, along with increased jetting velocities near the poles comes dramatically increasing viscous stresses there, and hence more work is done on the fluid. Since $J(\zeta)\sim e\sqrt{1-\zeta^2}$ for $e=b/a \ll 1$, the optimal profile shown in Fig.~\ref{Figure8}a requires a much more significant placement of jetting nozzles near the fore and aft poles, with a slight reduction of flow strength at $\zeta=\pm 1$. 

The optimal efficiencies for a range of aspect ratios are shown in Fig.~\ref{Figure9}a as filled circles, along with the exact efficiency calculated for the particular jetting profile $\phi(\zeta)=q\zeta/[\pi J(\zeta)]$ from Eq.~\eqref{E:Epro} as a dashed line.  (The other plotted values correspond to partially porous bodies, as discussed in the following section). While the optimal profile differs from that as studied analytically in the previous section, the efficiency increases only very slightly for each aspect ratio, by approximately $10^{-3}$. The integrated form of the efficiency measure appears to be insensitive to the precise form of the jetting flow profile for prolate jetting bodies, provided the general behavior of approximate geometry-normalized linear growth in $\zeta$. 

\subsection{Efficiency and optimization of a partially porous body}
\begin{figure}[t]
\begin{center}
\includegraphics[width=5.2in]{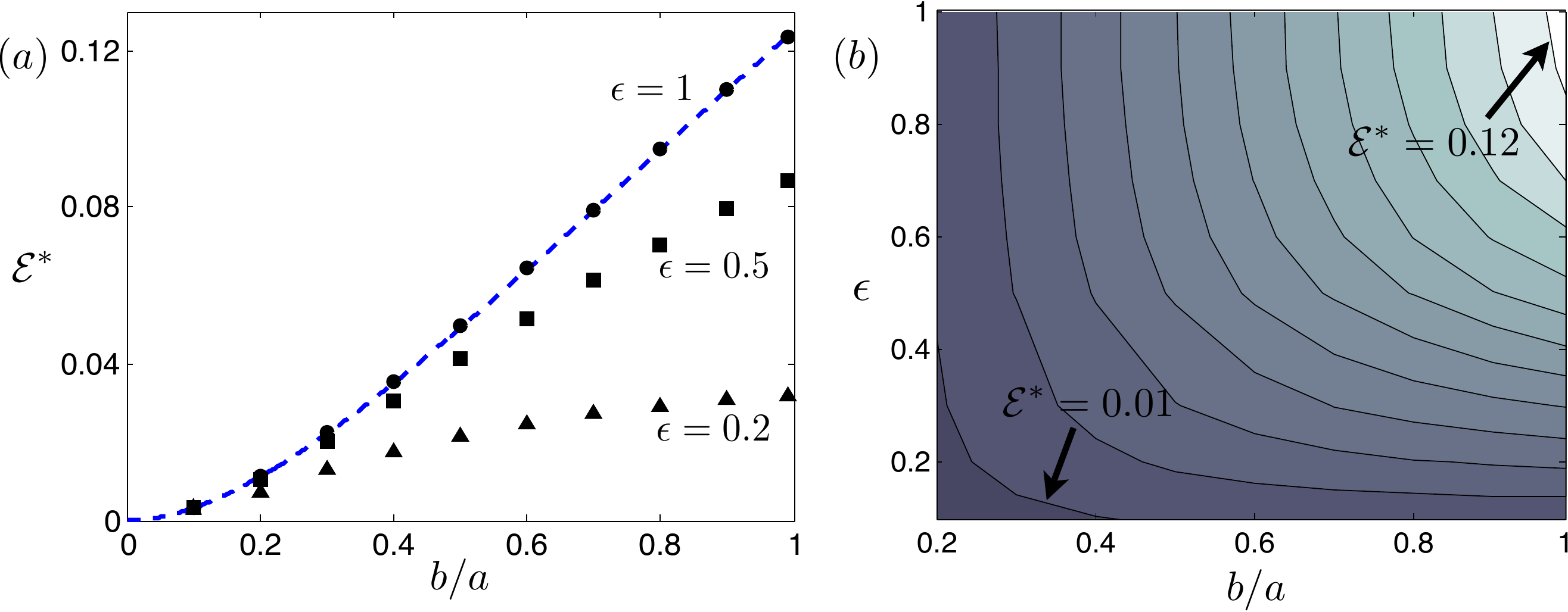}
\caption{(a) Optimal swimming efficiency as a function of aspect ratio for a selection of nozzle sizes, limiting to $\mathcal{E}^*=1/8$ as the body becomes spherical and entirely porous. The efficiency given by the jetting profile from the previous section $(\phi(\zeta)=q\zeta/(\pi J(\zeta)),\, \e=1)$ is shown as a dashed line. (b) Maximum efficiency for a given aspect ratio $b/a$ and nozzle size $\e$. Contours are shown at efficiency value multiples of $0.01$.}
\label{Figure9}
\end{center}
\end{figure}

Just as in the spherical case, we can ask about the optimal jetting profile and maximal efficiency when the swimming body's surface is only partially porous. Here we assume that the body is porous near the fore and aft poles in the regions $|\zeta|>1-\e$ (or $|z|>(1-\e) a$), and satisfies a no-slip condition along $|\zeta| \leq 1-\e$. The parameter $\e$ is the analogue of the spherical cap height parameter from Fig.~\ref{Figure4}, and the vertical measure of the porous cap height is again $\e\, a$. The optimization is performed numerically as before, but including the constraint that $\phi(\zeta)=0$ for $|\zeta|<1-\e$.

The maximum efficiency, $\mathcal{E}^*$, is shown for a range of aspect ratios $b/a$ and dimensionless cap-lengths $\e$ in Figs.~\ref{Figure9}a-b. For a given $\e$ the efficiency decreases as the body becomes more slender for the reasons already discussed. For a given aspect ratio, the maximal efficiency must not increase as the body becomes less porous, since the optimal profile for a more porous body could have $\phi(\zeta)=0$ wherever necessary. The efficiency becomes less sensitive to the porous surface size as the body becomes more prolate in shape, even though the jetting profile changes rather significantly as $\e$ is decreased.  Compared to the efficiencies of many other propulsive mechanisms at low Reynolds number, including those exploited by nature, the fluid jetting body compares favorably for all but the most slender of bodies, or the smallest of porous cap lengths. 

\begin{figure}[t]
\begin{center}
\includegraphics[width=3.6in]{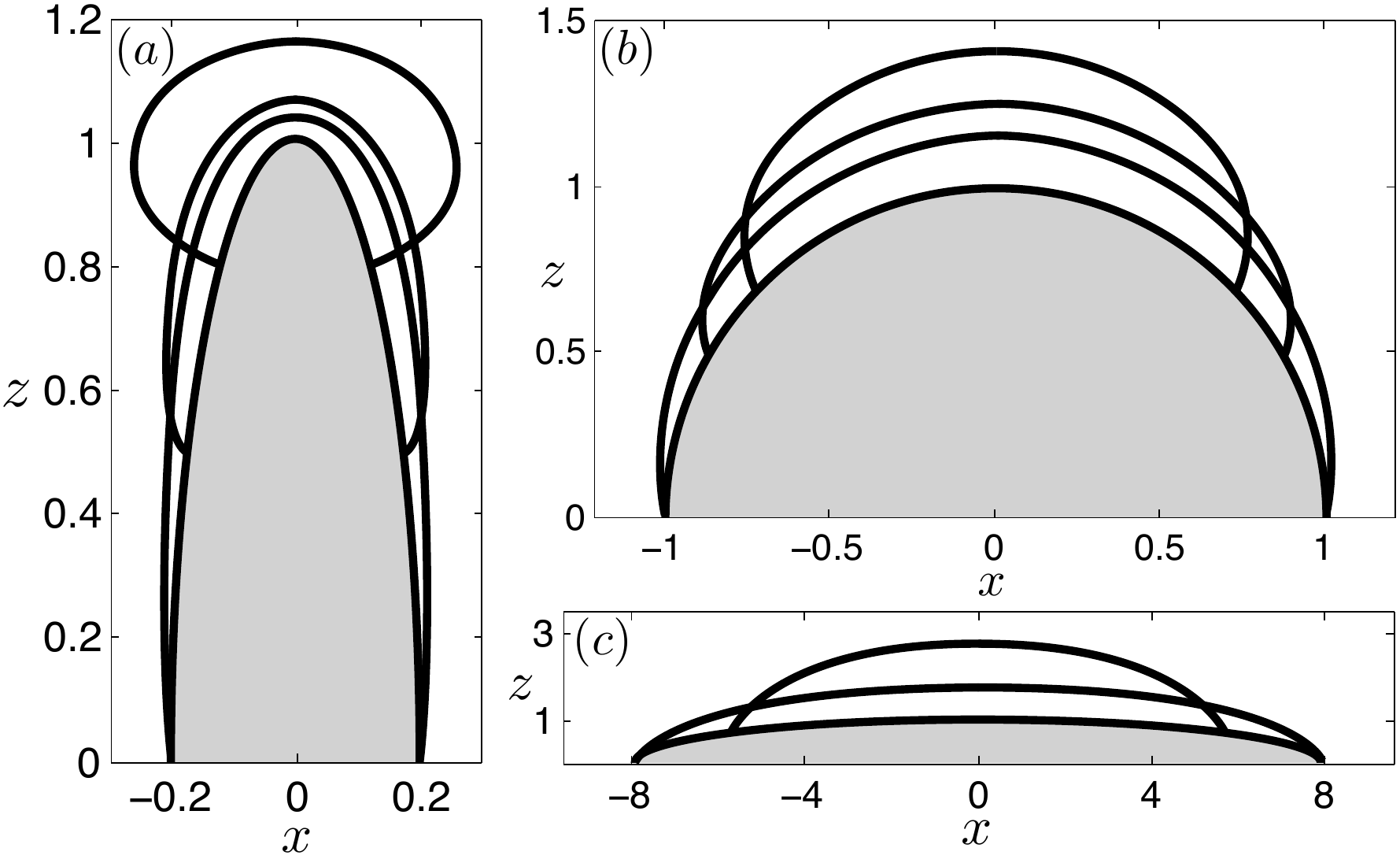}
\caption{The optimal jetting profiles are illustrated for a (a) prolate spheroid with $b/a=0.2$, (b) sphere, and (c) oblate spheroid with $a/b=8$. For each value of $\e$, the distance of the curve from the body surface corresponds to the jetting flow speed there. (The distance is scaled differently in each case for clarity of presentation.)}
\label{Figure10}
\end{center}
\end{figure}

The optimal jetting profiles are illustrated for a body with $b/a=0.2$ for $\e=1$, $\e=0.5$, and $\e=0.2$ in Fig.~\ref{Figure10}a. For each value of $\e$, the distance of the curve from the body surface corresponds to the jetting flow speed there. (The distance is scaled differently in each case for clarity of presentation.) The profile varies slowly in $\zeta$ in the fully porous case, and much more dramatically in the $\e=0.2$ case. However, all three profiles correspond very nearly to $\mathcal{E}\approx 1\%$ as indicated in Fig.~\ref{Figure9}a.  


\section{Oblate Body Shape}

Generally, efficient swimming in viscous fluids requires a streamlined body shape, and drag anisotropy is frequently utilized for propulsion. We have seen in the previous section that the hydrodynamic efficiency of a jetting body decreases as the body shape becomes more prolate, due to the reduction in the presented surface area in the direction of motion. The jet nozzles becomes less aligned with the swimming direction, and only the nozzles near the poles may be used for propulsion, adding significantly to the hydrodynamic work done on the fluid in the process. We now explore bodies in the shape of  oblate spheroids, and determine the consequences on swimming efficiency. We proceed just as in \S IV, benefitting from a simple correspondence between the prolate spheroidal and oblate spheroidal coordinate systems.

\subsection{An exact solution}
\label{oblate_analytical}
The results from the previous section may be used to determine the swimming velocity and efficiency for an oblate spheroid which is moving along the direction of its symmetric axis. The prolate spheroidal coordinate system is transformed to an oblate spheroidal coordinate system by replacing $c$ by $i c$, $\tau$ by $-i \lambda$, and reversing the roles of $a$ and $b$ so that $a>b$ in both cases. The surface area element becomes $dS=\hat{J}(\zeta)\,d\zeta\,d\theta$, with $\hat{J}(\zeta)=c^2 \sqrt{(\lambda_0^2+1)(\lambda_0^2+\zeta^2)}=a\sqrt{b^2+(a^2-b^2)\zeta^2}$.

The exact solution in the prolate case is simply converted to the oblate spherical coordinate system, and we assume the analogous jetting profile $\phi(\zeta)=q \zeta/[\pi \hat{J}(\zeta)]$. The swimming velocity is then
\begin{gather}
U=\frac{-q [1-\lambda_0\cot^{-1}(\lambda_0)]}{c^2 \pi }\cdot
\end{gather}
As the body becomes spherical ($\lambda_0\rightarrow \infty$) the swimming speed limits to the expected value $U=-q/(3\pi a^2)$. If instead we consider the limiting case of a flat-plate moving along its axis of symmetry ($b\rightarrow a$, $\lambda_0\rightarrow 0$), we find that the swimming speed becomes $U=-q/(\pi a^2)$. This result may not be surprising, given that as $\lambda_0 \rightarrow 0$ the jetting profile becomes $\phi(\zeta)=q/(\pi a^2)H(\zeta)$, where $H(\zeta)$ is the Heaviside function with a jump at $\zeta=0$. The flow speed is constant everywhere on the body surface outside the disk edge $\zeta=0$, and the swimming speed is precisely that opposing this jetting flow speed. 

The efficiency of an oblate spheroidal jetting body with this jetting profile becomes, transforming Eq.~\eqref{E:Epro},
\begin{gather}
\mathcal{E}=\frac{2 \left(\lambda_0^2+1\right) [\lambda_0 \cot^{-1}(\lambda_0)-1]^2}{\left[\left(\lambda_0^2-1\right)\cot^{-1}(\lambda_0)-\lambda_0\right]^2}\cdot\label{E:Eob}
\end{gather}
In the spherical limit the efficiency becomes $\mathcal{E}=12.5\%$ as expected, and as the body becomes a flat plate we find
\begin{gather}
\lim_{\lambda_0\rightarrow 0}\mathcal{E} = \frac{8}{\pi^2}\approx 81\%.
\end{gather}
Even without further optimization of the jetting velocity profile, the jetting body outperforms such ubiquitous propulsive mechanisms as flagellar undulations by an order of magnitude in this measure. 

\begin{figure}[t]
\begin{center}
\includegraphics[width=5.2in]{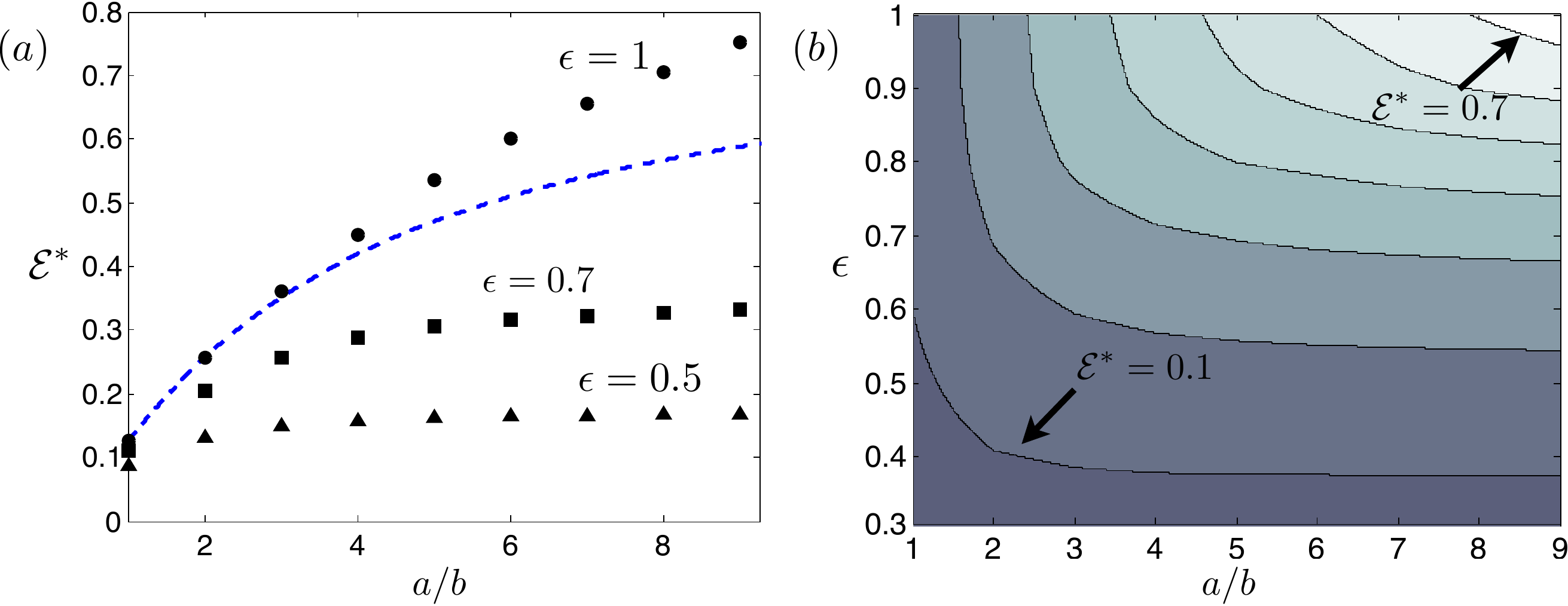}
\caption{(a) Optimal swimming efficiency as a function of aspect ratio for a selection of nozzle sizes, limiting to $\mathcal{E}^*=1/8$ as the body becomes spherical and entirely porous. The efficiency given by the jetting profile from the previous section $(\phi(\zeta)=q\zeta/(\pi J(\zeta)),\, \e=1)$ is shown as a dashed line. (b) Maximum efficiency for a given aspect ratio $b/a$ and nozzle size $\e$. Contours are shown at efficiency value multiples of $0.1$.}
\label{Figure11}
\end{center}
\end{figure}

\subsection{Efficiency and optimization of an oblate jetting body}

The jetting profile is now optimized numerically as in the previous section. Figure~\ref{Figure11}a shows the maximal efficiency found as a function of the aspect ratio $a/b$ for an entirely porous body ($\e=1$), along with the efficiency from the exact values from the previous section for the particular jetting profile considered there (as a dashed line). Unlike for prolate bodies, here there is a dramatic increase in the efficiency for very oblate bodies upon optimization. The jetting profile used to find an exact solution in the previous section limits to a Heaviside function as the body becomes a flat-plate. A consequence of this dramatic jump in the surface jetting velocity is a high wavenumber dissipation, with its associated deleterious effect on the swimming efficiency. Perhaps, then, the optimal jetting profile should have a tapered magnitude near the curve $\zeta=0$. This suspicion is confirmed numerically, as shown in Fig.~\ref{Figure8}b for a jetting body with aspect ratio $a/b=8$. Outside a region near $\zeta=0$ the optimal profile (normalized by $\hat{J}(\zeta)/q$) is approximately linear in $\zeta$. 

\begin{figure}[t]
\begin{center}
\includegraphics[width=3.0in]{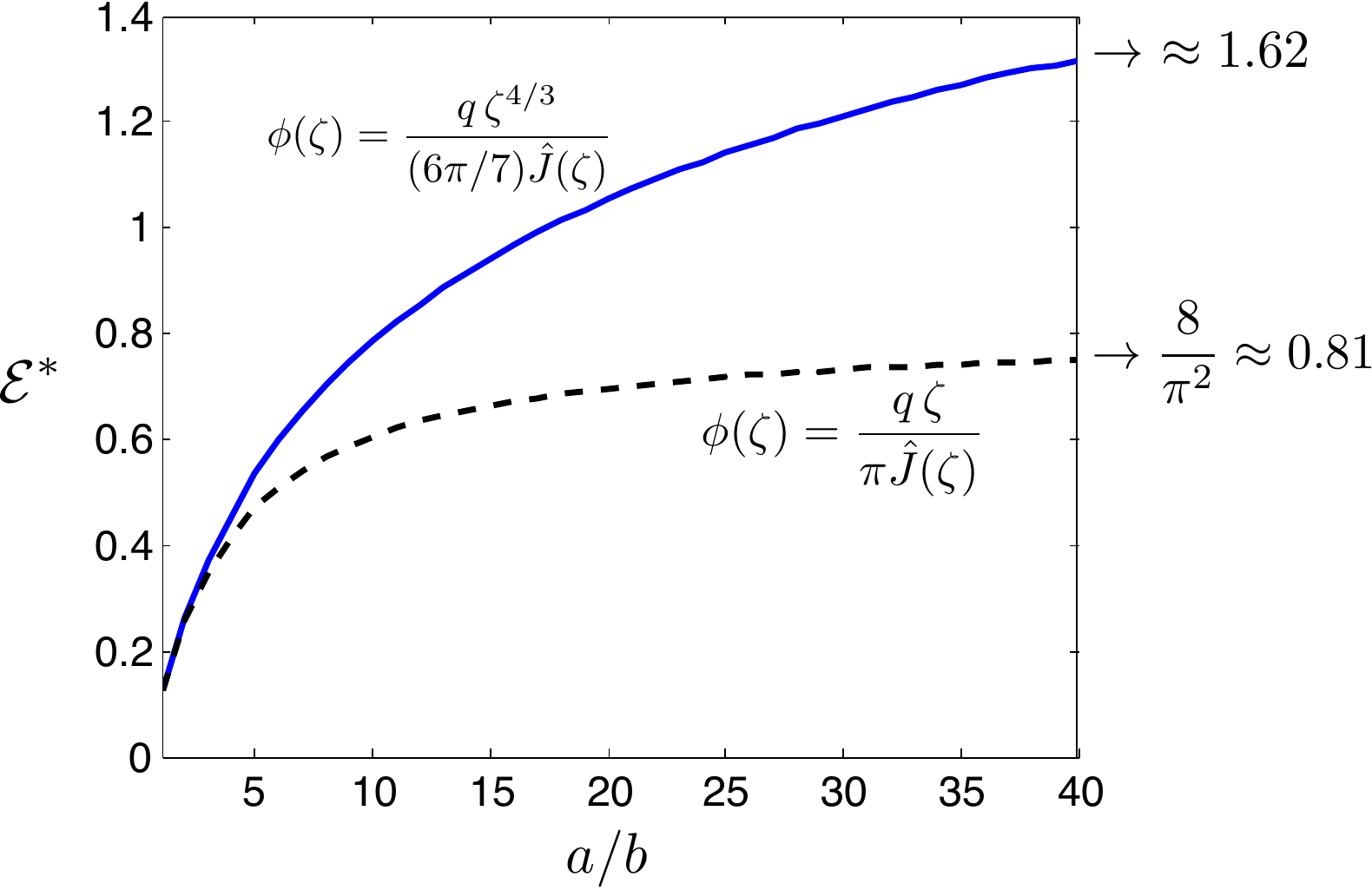}
\caption{The efficiency corresponding to two assumed jetting profiles for a large range of aspect ratios. Dashed line: analytical profile studied in \S\ref{oblate_analytical}. Solid line: approximate jetting profile leading to the maximum swimming efficiency (Eq.~\ref{E:newguy}), which yields an efficiency tending towards approximately $\mathcal{E}\approx162\%$ as the body becomes a flat plate.}
\label{Figure12}
\end{center}
\end{figure}

Numerically, we find that the computational results for increasingly oblate bodies falls increasingly close to the approximate jetting profile
\begin{gather}
\phi(\zeta)\approx \frac{q\,\zeta^{4/3}}{(6\pi/7)\hat{J}(\zeta)}\cdot\label{E:newguy}
\end{gather}
We may insert Eq.~\eqref{E:newguy} into the framework developed above and determine the corresponding swimming efficiency for any range of aspect ratios. The result is shown in Fig.~\ref{Figure12}. The efficiency corresponding to the above jetting profile is significantly larger than that of the previously considered analytical case, appearing to slowly limit as the body becomes a flat-plate to a value of $\mathcal{E}\approx162\%$, which is well above one. This value is also confirmed by our optimization results. For oblate-shaped bodies, the swimming efficiency is therefore quite sensitive to the (geometry-normalized) form of the jetting profile.

Finally, the partially porous case is considered for oblate spheroidal bodies. The maximal efficiencies are shown in Figs.~\ref{Figure11}a-b for a range of aspect ratios $a/b$ and cap distances $\e$ (where the surface is porous again on $|\zeta|>1-\e$, or $|z|>(1-\e) b$). The optimal profiles are illustrated in Fig.~\ref{Figure10}c for $\e=1$ (entirely porous) and $\e=0.5$. As in the previous considerations the efficiency can only decrease with decreasing $\e$.  However, for a given porous surface area we find that the efficiency does not vary dramatically as the body becomes more oblate. When the entire surface is porous we have already seen that tapering the jet strength near $\zeta=0$ can have a sizable impact on the swimming efficiency. If the body is only partially porous, the jetting profile near the edges is already zero, and the primary consequence of increasing the oblate aspect ratio must be a variation in the work done in the resistance problem. Since the drag on an oblate spheroidal body changes very little with the body aspect ratio, the slowly varying efficiency profiles might not be particularly surprising.

\section{Discussion}

\subsection{Summary of our results}
In this paper we have shown that a body can swim with remarkable efficiency by drawing in and expelling fluid at zero Reynolds number using non-inertial jet propulsion. Optimization of the jetting velocity profile at the surface yielded (external) swimming efficiencies of $\mathcal{E}=12.5\%$ for a fully-porous spherical body, and as much as $\mathcal{E}\approx 162\%$ as the body becomes very oblate in shape. In comparison to other common mechanisms utilized at low Reynolds numbers such as flagellar undulations, non-inertial jet propulsion thus presents an improvement of two orders of magnitude \cite{bw77,lp09}. 

\subsection{A locomotion mechanism for all Reynolds numbers}
An interesting aspect of jet propulsion, unlike such mechanisms as an undulating flagellum, is that a fluid-jetting body can self-propel at all Reynolds numbers, and it does so by exploiting distinct physical regimes. For large Reynolds numbers the motion of bodies generated by the expulsion of fluid jets has generated a vast literature, both in the exploration of jet propulsion in nature and for engineering purposes (see for example Refs.~\cite{Weihs77,dg88,lt04,dg05,Mohseni06,Dabiri08}). Unlike at zero Reynolds number, where the swimming efficiency is independent of the fluid flux $q$, greater thrust and efficiency may be achieved at higher Reynolds numbers by tuning properly the vortex ring formation in a pulsatile jetting locomotion \cite{Dabiri08}. At low Reynolds numbers the non-inertial jet propulsion utilizes the generation of viscous stresses in the surrounding fluid for propulsion, while at higher Reynolds numbers the mechanism shifts to the transmission of momentum into the fluid opposite the body motion. In contrast, the locomotion of a rotating helix, for example, while effective in a Stokesian realm, is all but useless at higher Reynolds numbers.

\subsection{The locomotion of {\em Synechococcus}}

A strain of cyanobacteria, {\em Synechococcus}, swims in fluid, but does so without the use of flagella \cite{wwfvw85,esbm96,ss96,Brahamsha99}. In fact, their motility mechanism is still an open problem in biophysics. Stone and Samuel \cite{ss96} proposed a model where {\em Synechococcus} was assumed to locomote by compressive surface distortions, i.e. the passage of traveling waves, tangentially along the body surface. Here we have shown that another mechanism of motility, one which includes a surface flow which acts in the direction normal to the body, can also provide an efficient locomotion. The motility of slime-extruding organisms such as cyanobacteria and myxobacteria (see Fig.~\ref{Figure1}) appears experimentally to depend significantly on the non-Newtonian rheology of the extruded slime, and the presence of a substrate. However, {\em Synechococcus} is known to swim absent the presence of a substrate, without changing shape, and without any observable external organelles. 

Given the recent observations of porous fluid extrusion in the biological community \cite{Walsby68,gc82,hb98,Hoiczyk00,whko02,wo04}, the work presented here might suggest an alternative motility mechanism for {\em Synechococcus}, that of non-inertial jet propulsion. To this end, let us approximate the swimming speed and efficiency of {\em Synechococcus} under the assumption that it utilizes fluid extrusion as a propulsive mechanism. The organism shape is approximately that of a prolate spheroid with aspect ratio $b/a\approx 1/2$, with cell length $2a\approx 10^{-4} $ cm. Wolgemuth et al. \cite{whko02} note that a related organism has nozzles everywhere on the surface, but Fig.~\ref{Figure9} indicates that the efficiency is not exceedingly dependent on the porous surface size. For the wide range $0.2 \leq \e \leq 1$ we recover an optimal swimming efficiency of $2\%-5\%$, well within the range of efficiencies reached by other microorganisms, if not slightly larger. To compute the swimming speed we need the flux $q$. To approximate the value of $q$ we use the estimated figures of Wolgemuth et al. \cite{whko02} for a related organism {\it M. xanthus}. Namely, that there are approximately $N=500$ nozzles around the circumference of each end of the cell, each nozzle has cross-sectional area $A\approx 3\times10^{-13}$  cm$^2$, and the fluid exit velocity is $u\approx 10^{-5} $ cm/s. This yields a fluid flux at each end of the cell of $q= N u A\approx 1.5\times 10^{-15} $ cm$^3$/s. From Eq.~\eqref{E:Upro}, and assuming that jet propulsion is  indeed the main locomotion mechanism, the swimming velocity would be approximately $|U|\approx 0.2 q/a^2=1.2\times 10^{-7}$ cm/s $=.0012\, \mu $m/s. This swimming speed is many orders of magnitude smaller than the observed swimming speed of {\em Synechococcus}, which is closer to $10\, \mu$m/s. In order for the swimming velocity to match that of {\em Synechococcus}, the flux would have to be $q\approx 10^{-11}$ cm$^3$/s, which would require a volume equivalent to the internal body volume to be expelled in an exceedingly fast time, $t\approx 10^{-2} $s. We can therefore conclude that the expulsion of a Newtonian fluid is unlikely to be the propulsive mechanism utilized by {\em Synechococcus}, giving support to other theories such as the passage of traveling surface waves of tangential displacements and of small amplitude \cite{ss96}.

\subsection{Directions for future work}
Exciting directions which this line of inquiry may take include optimization and control for jetting motions in three dimensions. For example, for a body with a known generalized resistance matrix, how best to utilize a finite collection of surface jets to move from one point to another in space? Separately, interaction dynamics may yield unexpected dynamics for low Reynolds number swimming behavior given the possibility of drawing in fluid just expelled by other swimmers. Finally, in the physical realization of such a swimmer, the internal mechanisms  for driving the flow both into and out of the body must be addressed. A more detailed efficiency analysis would then be of great interest, where internal costs are considered in addition to the hydrodynamic work performed on the external fluid. 

\begin{figure}[t]
\begin{center}
\includegraphics[width=1.6in]{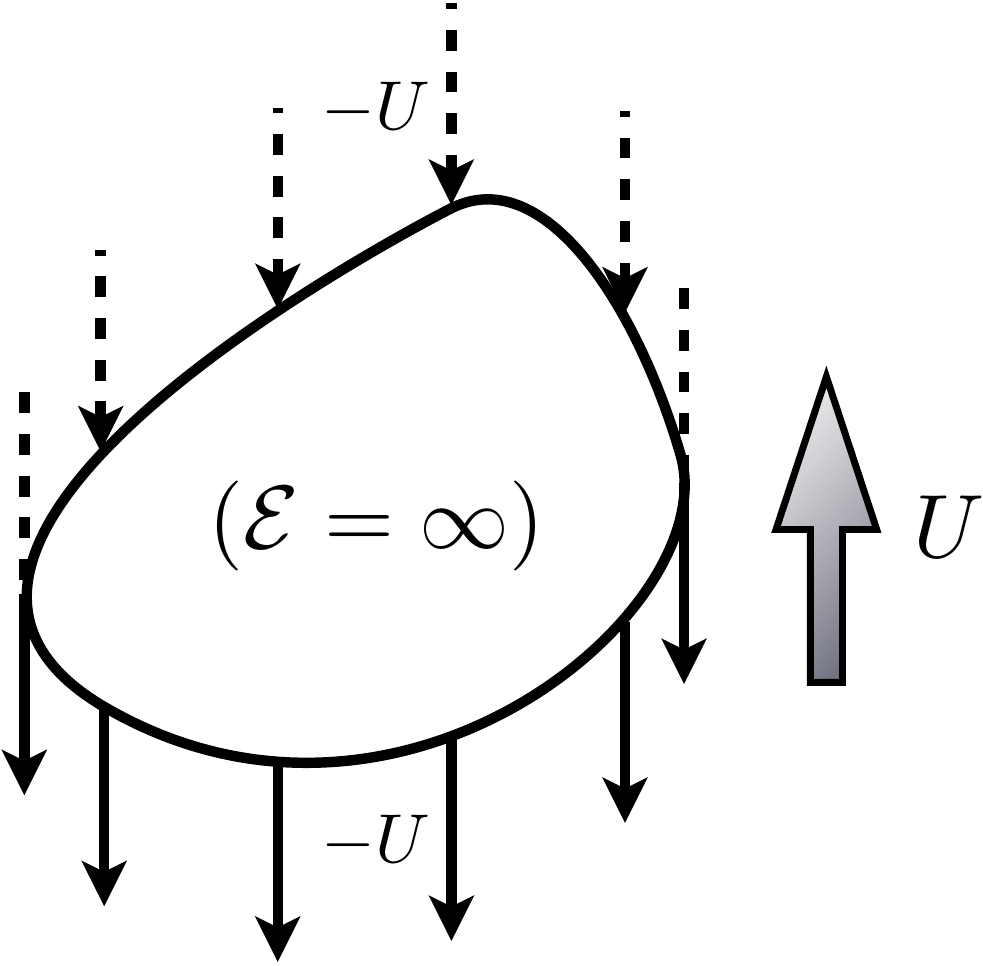}
\caption{An arbitrarily shaped body does not disturb the fluid and swims with infinite hydrodynamic efficiency if the surface distortion velocity is constant and unidirectional.}
\label{Figure13}
\end{center}
\end{figure}

We have seen that a fluid-jetting body can swim with remarkable efficiency. The constraint that fluid flow in and out along the direction everywhere normal to the body surface led to a bound on the efficiency of $12.5\%$ for a spherical body, and approximately 162\% as the body becomes a flat plate. Imagine this constraint to be removed. In this case the body could draw in and expel fluid everywhere upon its surface precisely in the direction of motion, and the surrounding fluid would go undisturbed (see Fig.~\ref{Figure13}). Hence the work done externally on the surrounding fluid would be zero, and the motion would correspond to an infinite hydrodynamic efficiency. Generally, the muted fluid disturbance even with the normal-jetting constraint hints at another very intriguing aspect of this form of locomotion: the body does not strongly signal its presence at any distance as it swims through the fluid. The non-inertial jetting body as portrayed in Fig.~\ref{Figure13} would be the perfect stealth swimmer.

\section*{Acknowledgements}
We thank Arthur Evans for fruitful conversations and gratefully acknowledge the support of the National Science Foundation through grant CBET-0746285.

\appendix
\section{Discontinuous jetting profile}

A discontinuous jetting profile results in zero swimming efficiency. To see this, consider a spherical body of radius $a$ with the jetting profile 
\begin{gather}
\phi(\zeta)=\frac{q}{2\pi a^2}\left(\chi_{\{\zeta\in[0,1]\}}-\chi_{\{\zeta\in[-1,0]\}}\right),
\end{gather}
where $\chi$ is the indicator function. The swimming velocity is, from Eq.~\eqref{E:SphereU},
\begin{gather}
U=-\frac{1}{4\pi a^2}\int_{\partial D} \left(\b{\hat{z}\cdot\hat{n}}\right)\phi(\b{x})\,dS=-\frac{1}{2}\int_{-1}^1 \zeta\,\phi(\zeta)\,d\zeta=-\frac{q}{4\pi a^2}.
\end{gather}
Inner products against the Legendre polynomials give
\begin{gather}
c_n=\int_{-1}^1 \phi(\zeta)P_n(\zeta)\,d\zeta=\frac{(1-(-1)^n)q}{2\pi a^2}\int_{0}^1 P_n(\zeta)\,d\zeta=\frac{(1-(-1)^{n})q}{2\pi a^2 n(n+1)}\frac{d}{d\zeta} P_{n}(\zeta)\Big|_{\zeta=0},\label{E:cn}
\end{gather}
and we note the identities
\begin{gather}
\frac{d}{d\zeta}P_n(\zeta)\Big|_{\zeta=0}=n\,P_{n-1}(0),\\
P_n(0)=\frac{(-1)^{n/2}(n-1)!!}{n!!}\,\,\,\,(n\,\,\, \mbox{even}),
\end{gather}
with $n!!=n\cdot(n-2)\cdot...\cdot4\cdot2$. Since $P_n(0)\sim n^{-1/2}$ for $n\rightarrow \infty$, we have
\begin{gather}
c_n=\frac{(1-(-1)^{n})(-1)^{(n-1)/2}(n-2)!!}{(n+1)(n-1)!!}\left(\frac{q}{2\pi a^2}\right)\sim n^{-3/2}\,\,\,\, (n\rightarrow \infty).
\end{gather}
Thus, the sum in the denominator of Eq.~\eqref{E:SphereE} diverges. Taking the first $N$ Legendre modes to approximate $\phi(\zeta)$, the efficiency is made arbitrarily small, with $\mathcal{E}\rightarrow 0$ in the limit of the discontinuous jetting profile $\phi(\zeta)$. A stress singularity associated with the jump in surface velocity results in infinite work being done on the fluid.

\section{Framework for general solution for spheroidal bodies}

We present here the framework used to determine the behavior and efficiency of prolate spheroidal jetting bodies. The methodology for oblate spheroidal bodies is identical, under the mapping $(\tau,c)\rightarrow (-i \lambda, i\,c)$, and switching the roles of $a$ and $b$ (see Fig.~\ref{Figure8}).

The general solution to Eqs.~\eqref{E:pde}-\eqref{E:pde2} may be expressed using spheroidal harmonic functions $R_n$ and $S_n$, themselves expressed in terms of the Legendre polynomials of the first and second kind, $P_n$ and $Q_n$ respectively,
\begin{gather}
\psi_\delta=A_1\,R_1(\tau)R_1(\zeta)+\displaystyle\sum_{n\geq 1}\left[B_n+\tilde{B}_n\left(\tau^2+\zeta^2\right)\right]R_n(\zeta)S_n(\tau),\label{E:PsiProlate2},
\end{gather} 
(see \S IV). From the Legendre differential equation, we have $R'_n(x)=-n(n+1)P_n(x)$ and $S'_n(x)=-n(n+1)Q_n(x)$. For a decaying fluid velocity as $\tau\rightarrow \infty$ we must have $A_1=0$. If the jetting profile $\phi(\zeta)$ is assumed to be odd about $\zeta=0$, then we may set $B_n=\tilde{B}_n=0$ for all $n$ even.

Due to the exponential decay of the terms $Q_n(\tau_0)$ and $S_n(\tau_0)$ as $n\rightarrow \infty$, it is helpful to define the normalized coefficients
\begin{gather}
C_n=B_n\,n\,Q_n(\tau_0),\,\,\,\,\tilde{C}_n=\tilde{B}_n\,n\,Q_n(\tau_0),
\end{gather}
noting the tractable behavior of the ratio
\begin{gather}
\frac{S_n(\tau_0)}{n\,Q_n(\tau_0)}=\frac{Q_{n-1}(\tau_0)}{Q_n(\tau_0)}-\tau_0.
\end{gather}
Differentiating the expression Eq.~\eqref{E:PsiProlate2}, the boundary conditions (Eqs.~\eqref{E:pde2}) therefore yield
\begin{gather}
0=\sum_{n \geq 1}2\tau_0 \tilde{C}_nR_n(\zeta)\left[\frac{S_n(\tau_0)}{nQ_n(\tau_0)}\right]-(n+1)\left[C_n+\tilde{C}_n\left(\tau_0^2+\zeta^2\right)\right]R_n(\zeta),\\
J(\zeta)\phi(\zeta)=\sum_{n \geq 1}2\zeta\tilde{C}_nR_n(\zeta)\left[\frac{S_n(\tau_0)}{nQ_n(\tau_0)}\right]-(n+1)\left[C_n+\tilde{C}_n\left(\tau_0^2+\zeta^2\right)\right]P_n(\zeta)\left[\frac{S_n(\tau_0)}{Q_n(\tau_0)}\right].
\end{gather}
We determine the coefficients $C_n$ and $\tilde{C}_n$ numerically by enforcing the conditions above at a finite number of nodes in $\zeta\in[-1,1]$, corresponding to the Gaussian quadrature nodes of order $M$, and keeping the first $N$ terms in each summation. The resulting linear system is inverted using simple Gaussian elimination. 

With the coefficients $C_n$ and $\tilde{C}_n$ in hand (and hence $B_n$ and $\tilde{B}_n$), we may determine the pressure and the $\bm{\hat{\tau}\hat{\tau}}$ component of the rate-of-strain tensor $\b{E}$. Associated with the stream function in Eq.~\eqref{E:PsiProlate2}, Eq.~\eqref{E:Pressures2} gives
\begin{align}
\frac{\partial p}{\partial \zeta}=\sum_n&\frac{-2 n (n+1) \mu\,\tilde{B}_n}{c^3 \left(1-\zeta ^2\right) \left(\tau ^2-\zeta ^2\right)^2} \Big[2 \zeta  \left(1-\zeta ^2\right) P_n(\zeta) \left(n (n+1) \left(\tau^2-\zeta ^2\right) Q_n(\tau)+2 \tau S_n(\tau)\right)\\
&-R_n(\zeta) \left(\left(\zeta ^4+2 \tau ^2+3 \tau ^4+\zeta ^2 \left(2-8 \tau ^2\right)\right) Q_n(\tau)-2 \tau  \left(\tau^2-\zeta ^2\right) S_n(\tau)\right)\Big],
\end{align}
which is integrated to within machine precision accuracy using Gaussian quadrature. The integration constant may be set to zero since the pressure must be an odd function about $\zeta=0$. 
\begin{figure}[t]
\begin{center}
\includegraphics[width=2.7in]{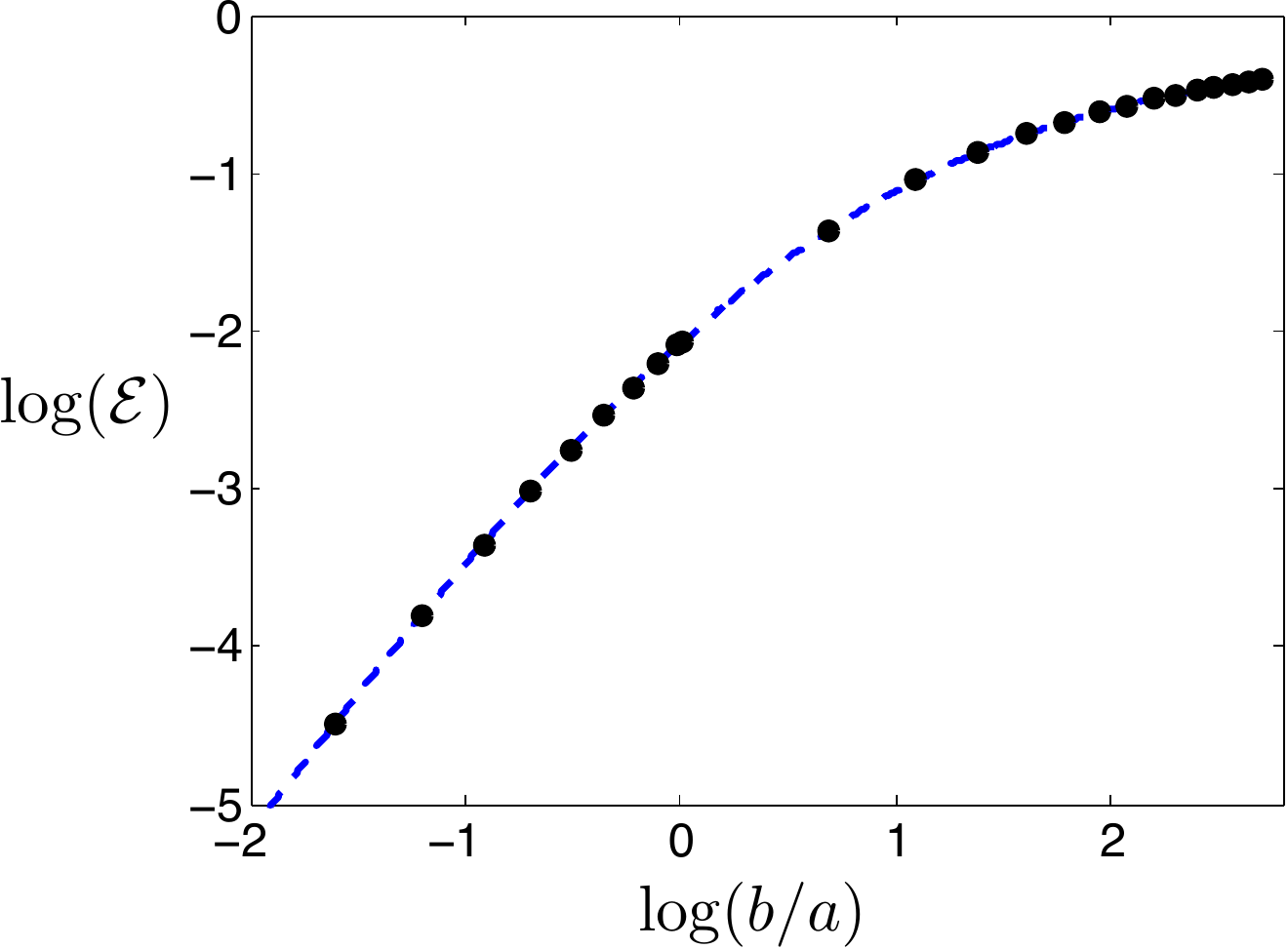}
\caption{Exact efficiency as a function of aspect ratio (with $a$ denoting the longer semi-axis length in both the prolate and oblate cases) is shown as a dashed line. The computed efficiencies using $M=100$ and $N=40$ are shown as filled circles, showing excellent agreement over a wide range of aspect ratios.}
\label{Figure14}
\end{center}
\end{figure}

From Eq.~\eqref{E:Etautau}, the desired rate-of-strain tensor component may be written as
\begin{eqnarray}
E_{\tau\tau}(\tau_0,\zeta)&=&\frac{1}{c^3(\tau_0^2-\zeta^2)^2(\tau_0^2-1)}\times\\
&&
\left[\tau_0(1+\zeta^2-2\tau_0^2)(\psi_\delta)_\zeta+\zeta(\tau_0^2-1)(\psi_\delta)_\tau+(\tau_0^2-\zeta^2)(\tau_0^2-1)(\psi_\delta)_{\tau\zeta} \right]\Big|_{\tau\rightarrow \tau_0}.
\nonumber
\end{eqnarray}
This component of the stress is dependent upon the jetting profile in a  straightforward manner. Namely, using the (arbitrary) jetting boundary conditions for $\psi_\delta$, we find that
\begin{gather}
E_{\tau\tau}(\tau_0,\zeta)=\left[\frac{\tau_0(1+\zeta^2-2\tau_0^2)}{c(\tau_0^2-\zeta^2)^{3/2}(\tau_0^2-1)^{1/2}}\right]\phi(\zeta).
\end{gather}
Finally, the efficiency is determined by numerically integrating the rate of work done on the fluid, 
\begin{gather}
\Phi=-2\pi\int_{-1}^1 \phi(\zeta)\left(-p+2\mu E_{\tau\tau}-p'\right)J(\zeta)\,d\zeta,
\end{gather}
and using the definition of Eq.~\eqref{E:Efficiency},
\begin{gather}
\mathcal{E}=\frac{6\pi\mu R U^2}{\Phi}\cdot
\end{gather}
The terms $p'$, $R$, and $U$ are just those as shown in Eqs.~\eqref{E:pPrime}, \eqref{E:Rdef}, and \eqref{E:UProlateG}, respectively.

The numerical approach is compared to the exact solutions of Eqs.~\eqref{E:Epro} and \eqref{E:Eob} in Fig.~\ref{Figure14}, using the jetting profile $\phi(\zeta)=\zeta q/[\pi J(\zeta)]$. Here we used $M=100$ and $N=40$. The computed efficiencies show excellent agreement with the analytical results over a wide range of body aspect ratios.

\bibliographystyle{unsrt}
\bibliography{BigBib}
\end{document}